\documentstyle[epsf,eqsecnum,floats,preprint,aps,epsfig]{revtex}

\input epsf
\tighten
\def\br{\mbox{\boldmath $r$}}
\def\bv{\mbox{\boldmath $v$}}

\def\high{\vphantom{\Biggl(}\displaystyle}

\begin{document}

\newcommand{\bm}[1]{\mbox{\boldmath{$#1$}}}
\newcommand{\be}{\begin{equation}}
\newcommand{\ee}{\end{equation}}
\newcommand{\bea}{\begin{eqnarray}}
\newcommand{\eea}{\end{eqnarray}}
\newcommand{\barr}{\begin{array}}
\newcommand{\earr}{\end{array}}

\rightline{CU-TP-1076}
\rightline{hep-th/0212328}
\vskip 1cm

\begin{center}
\ \\
\large{{\bf ADHMN boundary conditions from removing monopoles}} 
\ \\
\ \\
\ \\
\normalsize{Xingang Chen\footnote{email address:
xgchen@phys.columbia.edu} and Erick J. Weinberg\footnote{email
address: ejw@phys.columbia.edu} }
\ \\
\ \\
\small{\em Physics Department, Columbia University \\ New York, New
York 10027}

\end{center}

\begin{abstract}

Boundary conditions play an important role in the ADHMN construction
of BPS monopole solutions. In this paper we show how different types
of boundary conditions can be related to each other by removing
monopoles to spatial infinity.  In particular, we use this method to
show how the jumping data naturally emerge.  The results can be
interpreted in the D-brane picture and provide a better understanding
of the derivation of the ADHMN construction from D-branes. We comment 
briefly on the cases with non-Abelian unbroken symmetry and massless
monopoles. 

\end{abstract}

\setcounter{page}{0}
\thispagestyle{empty}
\maketitle

\eject

\vfill


\baselineskip=18pt

\section{introduction}
The Atiyah-Drinfeld-Hitchin-Manin-Nahm (ADHMN) construction
\cite{Atiyah:ri,Nahm:1979yw} is a powerful method for constructing the
Bogomol'nyi-Prasad-Sommerfield (BPS) \cite{BPS} magnetic monopole
solutions in Yang-Mills-Higgs theory. In this method, the problem of
solving the BPS equations for gauge and Higgs fields in
three-dimensional space is reduced to that of solving ordinary
differential equations for a triplet of Hermitian matrices $T_a$,
called the Nahm data, that are functions of a single variable $t$. We
will consider the case of an SU($N$) theory maximally broken to
[U(1)]$^{N-1}$ by an adjoint Higgs field with asymptotic form ${\rm
diag}(t_1,t_2,\dots,t_N)$, where $t_1<t_2<\cdots<t_N$ and $\sum t_i =
0$. There are $N-1$ species of ``fundamental monopoles''
\cite{Weinberg:1979zt}, each carrying a single unit of one of the
topological charges. The $T_a$ are defined on $t_1 \le t \le t_N$. On
the interval $t_i<t<t_{i+1}$, they are $(m_i\times m_i)$-dimensional,
where the $m_i$ give the numbers of the various fundamental
monopoles. We will denote this system as $(m_1, m_2,\dots, m_{N-1})$.

The boundary conditions are important in this construction. For
example, in SU(2) where we have only one interval $(t_1,t_2)$, the
Nahm data must have poles at the boundaries $t=t_1$ and $t=t_2$. For
SU($N$), where there are $N-1$ intervals, the boundary conditions at
$t_i$ depend on the numbers of fundamental monopoles on either side of
this boundary. When the numbers are unequal, the boundary conditions
are a generalization of those in SU(2) case.  There is an additional
element if $m_i=m_{i+1}$ for some $i$. In this case we also have
``jumping data'', consisting of a $2m_i$-component vector $a_i$
located at the boundary $t=t_i$.

While the constructions for the cases with or without jumping data are
different, the following intuitive physical picture suggests that
there must be a connection between them. Suppose we consider a
solution with $m_{i}>m_{i+1}$. We can gradually deform the solution by
removing the extra fundamental monopoles to spatial infinity, so that
the number of fundamental monopoles in $(t_i,t_{i+1})$ becomes equal
to $m_{i+1}$. We will study in this paper how the jumping data appear
in this process.

More generally, we will show how the ADHMN construction for magnetic
charge $(m_1,\dots,m_i,\dots,m_{N+1})$ goes over to a ``reduced''
ADHMN construction for charge $(m_1,\dots,m_i-1,\dots,m_{N+1})$ when
one of the fundamental monopoles is taken to spatial infinity.

There is a D-brane interpretation of this construction
\cite{Witten:1994tz,Douglas:1996uz,Diaconescu:1996rk,Kapustin:1998pb,Tsimpis:1998zh}.
In this picture the monopoles are D-strings ending on D3 branes
\cite{Callan:1997kz,Gibbons:1997xz}, $t$ parameterizes the direction
parallel to the D-strings, the $t_i$ are the places where the D3
branes are located and the Nahm data $T_a(t)$ describe the positions
and interactions of the various D-string segments. When the gauge
symmetry is maximally broken, this configuration can be thought of as
a superposition of different numbers of fundamental monopoles, with
each D-string joining adjacent D3 branes representing a fundamental
monopole.

The easiest way to get the jumping data is to start with a D0-D4
system and use T-duality to get the D1-D3 configuration
\cite{Kapustin:1998pb}. This requires that the numbers of D-string
segments (fundamental monopoles) on both sides of the D3-brane
(boundary) are the same. One finds that the jumping data originate
from the hypermultiplets that describe the D0-D4 strings. After the
T-duality, the hypermultiplets are localized at the positions of the
D3-branes (boundaries). Therefore, it is interesting to see how these
hypermultiplets (jumping data) appear or disappear when we remove
D-string segments (fundamental monopoles). This will help us to
understand how to consistently derive the other boundary conditions.

This paper is organized as follows: In Sec.~\ref{SectADHMN}, we
briefly review the ADHMN construction for SU($N$). In
Sec.~\ref{Sectsu2}, we study the case of SU(2). We show how the
$k$-monopole ADHMN construction becomes equivalent to the reduced
$(k-1)$-monopole construction when one monopole is removed to spatial
infinity. In Sec.~\ref{Sectpoles}, we study cases in larger groups
where no jumping data is involved. In Sec.~\ref{Sectjdata}, we
consider the cases $m_i=m_{i+1}+1$ and $m_i=m_{i+1}$, where jumping
data appears or disappears in the reduced problem.  An explicit SU(3)
example is given in Sec.~\ref{Secteg}. The results are interpreted in
terms of the D-brane picture in Sec.~\ref{Sectdpic}.
Section~\ref{SectConclu} contains the conclusions and some
discussion. Some details omitted from Sec.~\ref{Sectk>1} are
included in Appendix \ref{Appk>1}.

\section{The ADHMN construction}
\label{SectADHMN}
In this section, we briefly outline the ADHMN construction. We
restrict our discussion to the SU($N$) case \cite{Hurtubise:qy}. The
generalization to SO($N$) and Sp($N$) is discussed in
\cite{Hurtubise:qy}.  There are basically three steps in the ADHMN
construction. We first illustrate this in the SU(2) case with $m$
fundamental monopoles and asymptotic Higgs field ${\rm diag}(t_1,
t_2)$.

The first step is to find the Nahm data, which are $m \times m$
Hermitian matrices $T_a(t)$ $(a=1,2,3)$ defined on the interval $t_1
\le t \le t_2$. They satisfy the Nahm equations\footnote{Repeated
indices $a,b,c$ are summed over unless otherwise indicated. Here, and
for most of the paper, we set the gauge group coupling $e=1$.}
\begin{equation}
\frac{dT_a}{dt}= -i\epsilon_{abc}T_b T_c~.
\label{NahmEqn}
\end{equation}
It is useful to note that the Nahm equation is preserved under a
unitary transformation $T_a \rightarrow U T_a U^\dagger$, where $U$ is
$t$ independent.  The boundary conditions are that the $T_a$ have
simple poles at $t_1$ and $t_2$, with the residues being
$m$-dimensional irreducible SU(2) representations.

The second step is to solve the construction equation
\begin{equation}
0=\left[ -\frac{d}{dt} + (-T_a + r_a)\otimes\sigma_a
\right]v_{p}(t,r_a) ~,
\label{ConsEqn}
\end{equation}
where the $\sigma_a$ are the Pauli matrices, the $v_{p}$ are
$2m$-component vectors and $p$ labels the linearly independent
solutions. We note that the $v_p$ depend on the spatial position
$\br$, while the Nahm data $T_a$ do not; this $\br$ dependence will
often not be explicitly indicated. As we will see shortly, in the
SU(2) case there are two linearly independent $v_p$. We normalize
these so that 
\begin{equation} \int dt v_{p}^\dagger v_{q} = \delta_{pq}~.
\label{su2norm}
\end{equation}

In the third step we obtain the monopole gauge and Higgs fields
satisfying the self-dual BPS equations. If we assemble all the
independent normalizable solutions into a $2m_i \times 2$ matrix
$\bv$, then
\begin{eqnarray}
\Phi &=& \int dt \, t {\bf v}^\dagger {\bf v} ~, \nonumber\\ 
{\bf A}&=& -i \int dt {\bf v}^\dagger \nabla {\bf v} ~.
\label{su2fields}
\end{eqnarray}

The linearly independent solutions of the construction equation,
Eq.~(\ref{ConsEqn}), can be counted as follows: At each boundary, due to
the irreducible SU(2)-valued residue of the Nahm data, one can show
\cite{Hitchin:ay} that of the $2m$ solutions near each boundary, $m+1$
behave as $|t-t_{end}|^{(m-1)/2}$, while the other $m-1$ behave as
$|t-t_{end}|^{-(m+1)/2}$ and are thus non-normalizable. Matching the
$m+1$ normalizable solutions from the left boundary and the $m+1$
normalizable solutions from the right in the middle of the interval
imposes $2m$ constraints, because these vectors are
$2m$-dimensional. This leaves two independent normalizable solutions
and thus give the SU(2) fields in Eq.~(\ref{su2fields}). (Although the
case $m=1$ has no poles, the counting is the same.)

For the general SU($N$) case with asymptotic Higgs field ${\rm
diag}(t_1, t_2,..., t_N)$, the gauge symmetry is broken to
[U(1)]$^{N-1}$. Each U(1) factor is associated with a fundamental
monopole that can be obtained by embedding the unit SU(2) monopole.

On the interval $(t_i,t_{i+1})$ the Nahm data are $m_i\times m_i$
matrices $T_a^i$ obeying Eq.~(\ref{NahmEqn}). These define a
construction equation for $2m_i$-component vectors $v_{p}^i$. The
boundary conditions at $t_1$ and $t_N$ are the same as for SU(2). At
the other boundaries, the boundary conditions depend on the number of
fundamental monopoles on either side of the boundary. 

The $m_i \neq m_{i+1}$ case is a generalization of the SU(2) case. We
first assume $m_i \equiv m_{i+1}+k > m_{i+1}$.  Near the boundary
$t_{i+1}^-$, the $T_a^i$ take the form
\begin{equation}
\left( \begin{array}{cc}
S_a & {\cal{O}}\left( (t-t_{i+1})^{(k-1)/2} \right)\\ \\
{\cal{O}}\left( (t-t_{i+1})^{(k-1)/2} \right) & \high 
{-\frac{J_a^{(k)}}{t-t_{i+1}}}
\end{array}  \right),
\label{MatchingCond}
\end{equation}
where $S_a(t_{i+1})=T_a^{i+1}(t_{i+1})$, and the $J_a^{(k)}$ are
$k$-dimensional irreducible representations of SU(2). For the
solutions of Eq.~(\ref{ConsEqn}), the upper $2m_{i+1}$ components of
$v_p^i$ from $(t_i,t_{i+1})$ are continuous across the boundary,
connecting with the $2m_{i+1}$-component solutions from
$(t_{i+1},t_{i+2})$. The other $2k$ components of $v_p^i$ from
$(t_i,t_{i+1})$ are finite and terminate at the boundary. The case
$m_i<m_{i+1}$ is completely analogous.

In the case of $m_i=m_{i+1}$, the $T_a^i$ are discontinuous at the
boundary $t_{i+1}$. These discontinuities are described by an extra
term involving $2m_{i+1}$-dimensional row vectors $a^{i+1}_{r\alpha}$,
where $\alpha=1,2$ are spinor indices and $r=1,...,m_{i+1}$: 
\begin{eqnarray}
(\Delta T_a)_{rs} &\equiv&
T_a^{i+1}(t_{i+1})_{rs}-T_a^{i}(t_{i+1})_{rs} \nonumber \\ &=&
-\frac{1}{2} a_{s\alpha} (\sigma_a)_{\alpha\beta} a^{\dagger}_{r\beta}
\nonumber \\ &=& -\frac{1}{2} tr_2(a^\dagger_r a_s \sigma_a) ~.
\label{NahmEqnJump}
\end{eqnarray} [For simplicity we have dropped the superscript $(i+1)$ on $a$.]
The trace in the last equality is over the two-dimensional spinor
indices $\alpha$ of $a$ and the Pauli matrices $\sigma_a$.
Correspondingly, the solutions of the construction equations are also
discontinuous at the boundary, with
\begin{equation}
\Delta v = v^{i+1}(t_{i+1})-v^{i}(t_{i+1}) = - a^{(i+1)\dagger}
S^{i+1} ~, 
\label{ConsEqnJump}
\end{equation}
where the $S^{i+1}$ are complex numbers.

We can count the number of linearly independent normalizable solutions
of the construction equations by a method similar to that for the
SU(2) case. In the $k=0$ case, it is important to note that there is
an additional degree of freedom from the vector $a^{i\dagger} S^i$ in
Eq.~(\ref{ConsEqnJump}) when we connect the solutions from both side
of the boundary.  The final result is always $N$.

If we assemble the $S^i_p$ into an $N$-component row vector ${\bf
S}_i$, the normalization condition Eq.~(\ref{su2norm}) becomes
\begin{equation}
I=\int dt ~{\bf v}^\dagger {\bf v} +\sum_i {\bf S}_i^\dagger 
{\bf S}_i ~.
\label{NormCond}
\end{equation}
Equation~(\ref{su2fields}) becomes
\begin{eqnarray}
\Phi &=& \int dt~ t {\bf v}^\dagger {\bf v} + \sum_i t_i 
{\bf S}_i^\dagger {\bf S}_i \nonumber\\ 
{\bf A}&=& -i \int dt ~{\bf v}^\dagger \nabla {\bf v} - i\sum_i 
{\bf S}_i^\dagger \nabla {\bf S}_i~.
\label{Fields}
\end{eqnarray}
In these equations the sum over $i$ is restricted to the boundaries with
$m_{i-1}=m_i$.

\section{The SU(2) example}
\label{Sectsu2}
We first study the simplest example, that of SU(2) broken to U(1),
which does not involve the appearance or disappearance of the jumping
data. Parameter counting and other analyses \cite{Weinberg:ma} suggest
that in this case if all the monopoles are separated much further than
their core sizes, the solution can be approximated as a superposition
of many unit monopoles.

Let us assume that we have $k$ unit monopoles.  The Nahm data for this
system are $k \times k$ Hermitian matrices $T_a$ on the interval
$(t_1, t_2)$. We want to show that by removing one unit monopole, the
$k$ dimensional ADHMN construction effectively becomes that for $k-1$
monopoles. Doing this is also an explicit demonstration of the above
mentioned superposition picture. We assume that $k-1$ of the
monopoles, as well as the position $\br$ where we probe the fields,
are located within a region of size $l$, and that the $k$th monopole
is removed by a distance $D\gg l$, which without loss of generality we
can take to be along the $z$-axis.

The Nahm data have poles near the boundaries. This requires
\begin{equation}
T_a \approx -\frac{J_a^{(k)}}{t-t_2} 
\end{equation}
in the region $(t_2-1/D,t_2)$, and
\begin{equation}
T_a \approx -\frac{ {\tilde J}_a^{(k)} } {t-t_1} 
\end{equation}
in the region $(t_1,t_1+1/D)$,
where $J_a^{(k)}$ and ${\tilde J}_a^{(k)}$ are
$k$-dimensional irreducible representation\footnote{The $J_a^{(k)}$
and 
${\tilde J}_a^{(k)}$ do not have to be the same representation,
although they will of course be unitarily equivalent.} of SU(2).

Away from the boundaries, moving one fundamental monopole faraway
makes one of the eigenvalues in the Nahm data $T_a$ much larger than
the others. Therefore the Nahm data in the middle of the interval can
be put into the form
\begin{equation}
T_a=\left( \begin{array}{cc} M_a&A_a^\dagger\\A_a&b_a \end{array}
\right) ~, 
\label{TNotation}
\end{equation}
where the $M_a$ are $(k-1) \times (k-1)$ dimensional Hermitian
matrices with entries that are ${\cal O}(l)$, the $A_a^\dagger$ are
$k-1$ dimensional vectors that are at most ${\cal O}(\sqrt{lD})$, and
$b_a = \delta_{a3}D +{\cal O}(l)$.

It is useful to note that a unitary transformation $T_a \rightarrow U
T_a U^\dagger$ with
\begin{equation}
U=\left( \begin{array}{cc} I_{(k-1)\times(k-1)} + {\cal O}(l/D) &
-{\cal K}^\dagger/D \\ \\ {\cal K}/D & 1+{\cal O}(l/D)
\end{array} \right)
\label{utran}
\end{equation}
and ${\cal K} \le {\cal O}(\sqrt{lD})$ shifts
\begin{equation} A_a \rightarrow A_a - {\cal K} \delta_{a3} ~, ~~~
M_a \rightarrow M_a - \frac{{\cal K}^\dagger {\cal K}}{D} ~.
\label{Mshift}
\end{equation}
[In Eq.~(\ref{Mshift}) we have omitted terms that vanish when
$D\rightarrow \infty$.] By making use of such a transformation, we can
always subtract a $t$-independent constant from $A_3$. We will use
this freedom to make $A[(t_1+t_2)/2]$ vanish, up to exponentially
small terms.

In the following we will show that, in this case, the $A_a$ terms are
effectively negligible and that the $M_a$ obey the Nahm equations for
the $k-1$ monopole problem.  For $\br \sim {\cal O}(l)$, the fields
derived from $M_a$ using Eq.~(\ref{su2fields}) approximate those
derived from $T_a$.
 
We define
${A}=A_1+iA_2$, ${\hat A}=A_1-iA_2$, ${M}=M_1+iM_2$, 
${\hat M}=M_1-iM_2$, 
${b}=b_1+ib_2$ and $ {\hat b}=b_1-ib_2$.
The Nahm equations separate into the following equations:
\begin{eqnarray}
\frac{dM_a}{dt}&=&-i\epsilon_{abc}(M_b M_c + A_b^\dagger A_c) ~,
\label{EqnM} \\
\frac{d{A}}{dt}&=&-{A} M_3 + A_3 {M} - {b}A_3 +
b_3 {A} ~,
\label{Eqna} \\
\frac{d{\hat A}}{dt}&=&{\hat A} M_3 - A_3 {\hat M} + {\hat b}A_3 
- b_3 {\hat A} ~,
\label{Eqnah} \\
\frac{dA_3}{dt}&=&\frac{1}{2} \left( A {\hat M} - {\hat A}M + 
b{\hat A} - {\hat b}A \right) ~,
\label{EqnA3} \\
\frac{d{b}}{dt}&=&-{A}A_3^\dagger + A_3 {\hat A}^\dagger
~,
\label{Eqnb} \\
\frac{d{\hat b}}{dt}&=& {\hat A}A_3^\dagger - A_3 {A}^\dagger
~,
\label{Eqnbh} \\
\frac{db_3}{dt}&=& \frac{1}{2} \left( A A^\dagger - {\hat A}{\hat
A}^\dagger \right) ~.
\label{Eqnb3}
\end{eqnarray}

We first consider the middle of the interval, away from the
boundaries, where the $M_a$ are ${\cal O}(l)$.  Equation~(\ref{EqnA3})
gives
\begin{equation}
A_3={\cal O}(l)~|t_2-t_1|~{\rm Max}(A, \hat A) ~.
\label{A3order}
\end{equation}
(Recall that we have used a unitary transformation to subtract a
constant from $A_3$.)

If $A \gtrsim \hat A$, and\footnote{Since later we will know that $A_3
\sim e^{-D|t-t_2|}$ or $e^{-D|t-t_1|}$, this is an over-estimate. In
fact, we can replace $(t_2-t_1)$ by $1/D$; i.e., it is enough that $D
\gg l$.} $D \gg (t_2-t_1) {\cal O}(l^2)$, the fourth term dominates
the right-hand side of Eq.~(\ref{Eqna}). This gives
\begin{equation}
A = C e^{-D |t-t_2|} ~,
\label{Aform}
\end{equation}
where the coefficient $C$ is $t$-independent. Matching this
behavior to the pole region at $t \approx t_2-1/D$ requires that
$C={\cal O}(D)$. Consequently, Eq.~(\ref{EqnA3}) gives
\begin{equation}
A_3= C_3 e^{-D |t-t_2|} ~,
\label{A3form}
\end{equation}
where the coefficient $C_3$ is ${\cal O}(l)$. Equation~(\ref{Eqnah})
then implies 
\begin{equation}
{\hat A} = {\cal O}(l^2/D)~e^{-D|t-t_2|} ~.
\label{hAorder}
\end{equation}
If instead $\hat A \gtrsim A$,
\begin{eqnarray}
{\hat A} &=& {\hat C}e^{-D|t-t_1|} ~, 
\label{hAform2} \\
A_3 &=& C_3 e^{-D |t-t_1|} ~, 
\label{A3form2} \\
A &=& {\cal O}(l^2/D) ~e^{-D|t-t_1|} ~,
\label{Aorder2}
\end{eqnarray}
where $\hat C$ is ${\cal O}(D)$ and $C_3$ is ${\cal O}(l)$.
Equations~(\ref{Aform}-\ref{hAorder}) apply for $t$
closer to $t_2$, while Eqs.~(\ref{hAform2}-\ref{Aorder2}) apply for
$t$ closer to $t_1$.

{}From Eqs.~(\ref{Aform}-\ref{Aorder2}), we see that $A_a \ll {\cal
O}(l)$ except within narrow regions of width $t_D \equiv
\frac{1}{D}\ln\frac{D}{l}$ near the boundaries. Hence, outside these
boundary regions, Eq.~(\ref{EqnM}) can be approximated by the
$(k-1)$-monopole Nahm equations with Nahm data $M_a$. In order to
match the pole behavior of the $T_a$ 
in $(-1/D,0)$ in the original problem, the $(k-1)$-monopole Nahm data
$M_a$ 
must have pole behavior in $(-1/l,-t_D)$. The residues of the poles
will be irreducible SU(2) representations.

To see what happens to the construction equation, we decompose
\begin{equation}
v=\left 
( \begin{array}{c} w \\ z \end{array} 
\right) ~, 
\end{equation}
where $w$ is a $2m$-dimensional vector and $z$ is
2-dimensional. The construction equation then becomes
\begin{eqnarray}
0&=&-\frac{d}{dt}w + \left[ (-M_a + r_a) \otimes \sigma_a \right]
w - (A_a^\dagger \otimes \sigma_a)z ~,
\label{Eqnw} \\
0&=&-\frac{d}{dt}z - (A_a \otimes \sigma_a) w + \left[ (-b_a+r_a)
\otimes \sigma_a \right] z ~.
\label{Eqnz}
\end{eqnarray}

In the interval $(t_1+t_D, t_2-t_D)$ the $w$ and $z$ components are
decoupled, since the contributions to them from the cross terms in
Eqs.~(\ref{Eqnw}) and (\ref{Eqnz}) are negligible due to the
exponential smallness of the $A_a$. There are three types of
solutions. The first two types are associated with the reduced Nahm
data $M_a$. One is of the form
\begin{equation}
v=\left( \barr{c} v_i \\ 0 \earr \right)+\cdots~, \qquad i=1,2 
\label{type1}
\end{equation}
where $v_i$ is a normalizable solution of the $(k-1)$-monopole
construction equation formed from the $M_a$, and the dots represent
exponentially small terms. Next are solutions of the form
\begin{equation}
v=\left( \barr{c} u_j \\ 0 \earr \right)+\cdots~, \qquad j=1,\dots,2k-4
\label{type2}
\end{equation}
where the $u_j$ are non-normalizable solutions of the construction
equation formed from the $M_a$. These behave as $|t-t_{end}|^{-k/2}$
at least near one boundary. Finally, there are two solutions of
the form 
\begin{equation}
v=\left( \barr{c} 0 \\ z_{\pm} \earr \right) ~,
\label{type3}
\end{equation}
where $z_{\pm} = e^{\pm Dt} \eta_{\pm}$ and 
$\sigma_3 \eta_{\pm} = \mp \eta_{\pm}$, and higher order exponential
terms are ignored. As we can see, these are concentrated near the
boundaries.

We normalize the $v_i$ so that
\begin{eqnarray}
\int v_i^{\dagger} v_j &=& \delta_{ij} ~, 
\end{eqnarray}
and fix the scale in Eqs.~(\ref{type2}) and (\ref{type3}) by requiring
\begin{eqnarray}
{\rm Max} \left[ u_j(t_1+1/D), u_j(t_2-1/D) \right] &\sim& \sqrt{D} ~,
\\ 
{\rm Max} \left[ z_{\pm}(t_1+1/D),z_{\pm}(t_2-1/D) \right] &\sim&
\sqrt{D} ~. 
\end{eqnarray}
{}From the discussion below Eq.~(\ref{su2fields}), we see that
$v_i(t_1+1/D)$ and $v_i(t_2-1/D)$ are both of order $D^{-k/2+1}$.

Now we match the linear combination
\begin{equation}
v=\sum_{i=1}^{2} c_i v_i + \sum_{i=1}^{2k-4} d_i u_i + \sum_{i=\pm}
e_i z_i 
\label{vcombine}
\end{equation}
to the normalizable solutions in the boundary regions $(t_1,t_1+1/D)$
and $(t_2-1/D,t_2)$. As we have seen, within each of these two
boundary regions there are $k+1$ normalizable solutions behaving as
$|t-t_{end}|^{(k-1)/2}$.  The matching of the solutions of
Eq.~(\ref{vcombine}) on to the boundary region solutions requires that
all three terms in Eq.~(\ref{vcombine}) be of the same order of
magnitude at $|t-t_{end}|=1/D$. This implies that the $d_i$ and $e_i$
are order of $D^{-(k-1)/2}$ smaller than the $c_i$.  Therefore, the
last two terms of Eq.~(\ref{vcombine}) are negligible\footnote{Simply
stated, the solutions in Eqs.~(\ref{type2}) and (\ref{type3}) share the
feature that they decay quickly away from the boundaries, so if their
scales are fixed to be finite near the boundary, they do not
contribute to the fields. We will see similar situations later.}  in
the integrals in the normalization condition, Eq.~(\ref{su2norm}), and
the fields, Eq.~(\ref{su2fields}), since they only give corrections of
order $D^{-k+1}$.  The changes of the solutions of Eq.~(\ref{type1})
within $(t_1,t_1+t_D)$ and $(t_2-t_D,t_2)$ also only have negligible
effect on the fields and normalization condition, because the
integration region is too small. So we effectively recover the reduced
$(k-1)$ monopole construction.

\section{The pole behavior} 
\label{Sectpoles}
In this and the next section, we describe how the boundary conditions
in the ADHMN construction change when a monopole is removed to spatial
infinity. We will concentrate on the case of SU(3) with magnetic
charge $(m+k,m)$, focusing on the boundary between the intervals
$(t_1,t_2)$ and $(t_2,t_3)$; the extension to larger unitary groups is
straightforward.  We write the $(m+k)\times(m+k)$ Hermitian matrices
on the left of $t_2$ as $T^L_a$ and the $m\times m$ Hermitian matrices
on the right as $T^R_a$. Because we will remove one fundamental
monopole corresponding to the left interval $(t_1,t_2)$, one
eigenvalue of $T^L_a$ should be of order $D$ throughout most of this
interval. The other elements of $T^L_a$ and all the elements of
$T^R_a$ should be ${\cal O}(l)$ outside the boundary regions.  For
simplicity, we take $t_2=0$ in this and the next section.

We first study the two simpler cases where no jumping data is
involved. These are the case $k>1$ (in Sec.~\ref{Sectk>1}) and the
case $k<0$ (in Sec.~\ref{Sectk<0}). In the next section we study how
removing a monopole leads to the appearance (for $k=1$) and
disappearance (for $k=0$) of jumping data.

\subsection{$k>1$: Reducing the dimension of the pole term on the
left}
\label{Sectk>1}
We first consider the case where we remove one fundamental $(1,0)$
monopole from the $(m+k,m)$ system, with $k>0$.  (The discussion of
Sec.~\ref{Sectsu2} can be viewed as the special case where $m=0$.)

We first phrase the problem and our expectation. In the original
$(m+k,m)$ problem, we have on the left interval Hermitian matrices
$T^L_{a,{\rm orig}}$ that have one large eigenvalue in the middle of
the interval and that develop the poles in the boundary region
$(-1/D,0)$ of the form
\begin{eqnarray}
T_{a,{\rm orig}}^L= \left( \barr{cc} N_a & {\cal O}(t^{(k-1)/2}) \\ \\
{\cal O}(t^{(k-1)/2}) & \high
{-\frac{J^{(k)}}{t}} \earr \right)~,
\end{eqnarray}
where the $N_a(0) = T_a^R(0)$ are $m\times m$ dimensional. We want to
show that this is equivalent to the reduced $(m+k-1,m)$ problem whose
Nahm data $T^L_{a,{\rm red}}$ have poles in the boundary region
$(-1/l,0)$ of the form
\begin{eqnarray}
T_{a,{\rm red}}^L = \left( \barr{cc} N_a & {\cal O}(t^{(k-2)/2}) \\ \\
{\cal O}(t^{(k-2)/2}) & \high 
{-\frac{J^{(k-1)}}{t}} \earr \right)~.
\end{eqnarray}
Again, the $N_a(0) = T^R_a(0)$ are $m\times m$-dimensional.

We write the Nahm data in the left interval as
\begin{equation}
T^L_a=\left( \barr{c|c} M_a & A_a^\dagger \\ \hline A_a & b_a \earr
\right) 
= \left( \begin{array}{cc|c} N_a & E_a^\dagger & F_a^\dagger \\ 
E_a & P_a & G_a^\dagger \\ \hline F_a & G_a & b_a \end{array} \right)
~, 
\end{equation}
where in the second equality we have separated the
$(m+k-1)$-dimensional $M_a$ into
blocks of dimension $m$ and $(k-1)$.

\begin{figure}[htb]
\begin{center}
\epsfig{file=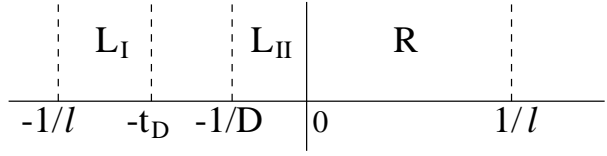, width=8cm}
\end{center} 
\medskip
\caption{The boundary region near $t=0$. We denote the interval
$(-1/l,-t_D)$ by ${\rm L_I}$, $(-1/D,0)$ by ${\rm L_{II}}$, and
$(0,1/l)$ by R.}
\label{tregion1}
\end{figure}

The same arguments as in the SU(2) case show that away from the
boundary, the $A_a$ and $A_a^\dagger$ are negligible and the $M_a$
obey the $(m+k-1)$-monopole Nahm equations.\footnote{As in the SU(2)
case, we can eliminate a constant from $A_3$ by a unitary
transformation of the form of Eq.~(\ref{utran}). We show in Appendix
\ref{Appk>1} that $A_3$ is at most of order $l$, so the shifts in the
$M_a$ are negligible.} However we must make sure that the $N_a$ are
constant in the boundary region $(-1/D,0)$ as $D\rightarrow \infty$,
so that they can continuously connect to the $T_a^R$ in the next
interval.  In Appendix \ref{Appk>1} we show that the transition of the
poles in the $T^L_a$ is similar to the SU(2) case. In terms of the
regions shown in Fig.\ref{tregion1}, we have
\begin{eqnarray}
{\rm Region~ L_{I} :}~~
T^L_{1,2}(t) &=& 
\left( \barr{cc} T^L_{1,2,{\rm red}} & {\cal O}(D)~e^{-D|t|} \\ \\
{\cal O}(D)~e^{-D|t|} & {\cal O}(l) \earr \right) ~, 
\qquad -1/l<t<-t_D ~,
\nonumber \\ \nonumber \\ 
T^L_{3}(t) &=& 
\left( \barr{cc} T^L_{3,{\rm red}} & {\cal O}(l)~e^{-D|t|} \\ \\ 
{\cal O}(l)~e^{-D|t|} & D + {\cal O}(l) \earr \right) ~, 
\qquad -1/l<t<-t_D ~,
\nonumber \\ \nonumber \\ 
{\rm Region~ L_{II} :}~
T^L_a(t) &=& T^L_{a,{\rm orig}} ~, \qquad -1/D<t<0 ~. \nonumber
\end{eqnarray}

The discussion of the construction equation is similar to the
SU(2) case and included in Appendix \ref{Appk>1}.

\subsection{$k<0$: Increasing the dimension of the pole term on the
right}
\label{Sectk<0}
We now remove a $(1,0)$ monopole
from an $(m-|k|,m)$ configuration and see how a higher dimensional
pole arises in $T^R_a$. In the original
$(m-|k|,m)$ problem, $T^L_{a,{\rm orig}}$ is $(m-|k|) \times
(m-|k|)$-dimensional. Near the boundary, $0<t<1/D$,
\begin{eqnarray}
T^R_{a,{\rm orig}} &=& \left( \barr{cc} N'_a & {\cal O}(t^{(k-1)/2})
\\ \\ {\cal O}(t^{(k-1)/2}) & 
\high{-\frac{J_a^{(|k|)}}{t}} \earr \right) ~,
\end{eqnarray}
where the $N'_a(0)=T_{a,{\rm orig}}^L(0)$ are $(m-|k|)\times
(m-|k|)$-dimensional. Here both $T^L_{3,{\rm orig}}$ and $N'_3$ have
one large eigenvalue $D$.
We want to show that this problem is equivalent to the reduced
$(m-|k|-1,m)$ problem where $T^L_{a,{\rm red}}$ is $(m-|k|-1)\times
(m-|k|-1)$-dimensional and ${\cal O}(l)$, while
\begin{eqnarray}
T^R_{a,{\rm red}} &=& \left( \barr{cc} N_a & {\cal O}(t^{k/2}) \\ \\
{\cal O}(t^{k/2}) & 
\high{-\frac{J_a^{(|k|+1)}}{t}} \earr \right) ~,
\end{eqnarray}
for $0<t<1/l$, with
$N_a(0)=T_{a,{\rm red}}^L(0)$.

We decompose the Nahm data $T^L_a$ as in Eq.~(\ref{TNotation}),
\begin{equation}
T^L_a=\left( \begin{array}{cc} M_a&A_a^\dagger\\A_a&b_a \end{array}
\right) ~, \nonumber
\end{equation}
but 
with the $M_a$ being $(m-|k|-1)$-dimensional. 

Because the off-diagonal terms $A_a$ must match on to the $N'_a$, they
can be at most of ${\cal O}(l)$ at $t=0$. This is different from the
situation in Sec.~\ref{Sectsu2}. Thus, in the large $D$ limit, the
only divergent element of $T^L_a$ at the boundary is $b_3=D$. Since
the $T^L_a$ are continuously connected to the $(m-|k|)\times (m-|k|)$
upper diagonal block of $T^R_a$, this requires that the corresponding
element of $T^R_a$ also be divergent at $t_2$. The Nahm
Eqs.~(\ref{NahmEqn}) require this divergent behavior to be a simple
pole and the residues to be $(|k|+1)$-dimensional SU(2)
representations, which is irreducible due to the irreducibility of the
residue in the original problem within $(0,1/D)$.

\begin{figure}[htb]
\begin{center}
\epsfig{file=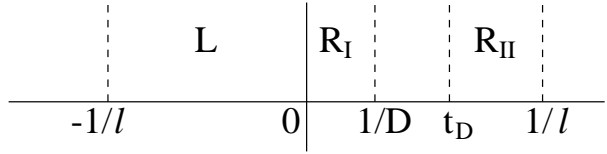, width=8cm}
\end{center} 
\medskip
\caption{The boundary region at $t=0$. We denote the interval $(-1/l,0)$
by L, $(0,1/D)$ by ${\rm R_I}$, and $(t_D,1/l)$ by ${\rm R_{II}}$.}
\label{tregion2}
\end{figure}

In terms of the regions shown in Fig.~\ref{tregion2}, the transition
of the pole behavior is\footnote{The block-diagonal form in region L
can be obtained by the unitary transformation Eq.~(\ref{utran}), as
before.  We now also do a corresponding unitary transformation $\left(
\barr{cc} U&0 \\ 0&I_{|k|\times|k|} \earr \right)$ on the $T_a^R$,
thus maintaining 
the boundary condition at $t=0$.}
\begin{eqnarray}
{\rm Region~L :}~~~
T^L_a(t) &=& \left( \begin{array}{cc} T^L_{a,{\rm red}} & {\cal
O}(l)~e^{-D|t|} \\ \\ {\cal O}(l)~e^{-D|t|} &
D\delta_{a3}+{\cal O}(l)\end{array} \right) ~, \qquad -1/l<t<0 ~,
\nonumber \\ \nonumber \\ 
{\rm Region~R_{I} :}~~
T^R_a(t) &=& T^R_{a,{\rm orig}} ~, \qquad 0<t<1/D ~, 
\nonumber \\ \nonumber \\ 
{\rm Region~R_{II} :}~
T_a^R(t) &=& T^R_{a,{\rm red}} ~, \qquad t_D<t<1/l ~. \nonumber
\end{eqnarray}

As in the previous case, the construction equation on the left
interval $(t_1,0)$ has a solution $v_+$ localized near $t=0$.  On the
right interval, there is a solution $v_R$ that behaves as $(\Delta
t)^{-(|k|+2)/2}$ in $(t_D,1/l)$ and that would have been discarded due
to non-normalizability in the $(m-|k|-1,m)$ description, that becomes
normalizable in $(0,1/D)$ because the dimension of the pole residue
has decreased. However, like $v_+$, this solution decays quickly away
from the boundary. By reasoning similar to that at the end of
Sec.~\ref{Sectsu2}, these solutions have no effect on the fields in
the $D\rightarrow \infty$ limit. Thus the original $(m+k,m)$
description goes over to the reduced $(m+k-1,m)$ description.

\section{The jumping data}
\label{Sectjdata}
\subsection{$k=1$: Emergence of the jumping data}
\label{Sectk=1}
As described in the introduction, we want to investigate how the
jumping data appear when a monopole is removed from a configuration
that originally has no jumping data. We start with an $(m+1,m)$
configuration in the SU(3) theory. The Nahm data of interest for this
system are three $(m+1) \times (m+1)$ Hermitian matrices $T_a^L$
defined on $(t_1,0)$, and three $m \times m$ Hermitian matrices
$T^R_a$ defined on $(0, t_3)$ that are continuously connected to the
upper diagonal $m \times m$ block of the $T^L_a$. We write
\begin{equation}
T^L_a=\left( \begin{array}{cc} M_a&A_a^\dagger\\A_a&b_a \end{array}
\right)
\label{TNotation1}
\end{equation}
where the $M_a$ are $m\times m$ dimensional, the $A_a^\dagger$ are
$m$-dimensional vector and the $b_a$ are real. Again, we choose
$b_a=\delta_{a3} D + {\cal O}(l)$. The Nahm equations are decomposed
as in Eqs.~(\ref{EqnM}-\ref{Eqnb3}).

It is easier to analyze the Nahm equation when they are
block-diagonalized in the middle of the interval. Using the same
arguments as in Sec.~\ref{Sectsu2}, we first do a unitary
transformation to subtract an additive constant from $A_3$. We will
denote the matrices after this transformation with an extra prime. Two
of the decomposed Nahm equation become
\begin{eqnarray}
\frac{dM_a'}{dt}&=&i\epsilon_{abc}(M_b' M_c' + {A_b'}^\dagger A_c')~,
\label{EqnM'} \\
\frac{dA_3'}{dt}&=&i\epsilon_{3bc}(A_b' M_c' + B_b' A_c') ~,
\label{EqnA3'} 
\end{eqnarray}

The exponential dependence of $A$ in Eq.~(\ref{Aform}) can now be
extended all the way to the boundary $t=0$. This is because, unlike
the SU(2) case in Sec.~\ref{Sectsu2}, we no longer have poles at the
boundary. The $M_a$ are required to continuously connect to the
matrices on the right interval and are always ${\cal O}(l)$. We then
have
\begin{equation}
A'_{1,2}=\tilde A_{1,2} \sqrt{2D}e^{-D|t|} ~,
\label{A'12order}
\end{equation}
where the $t$-independent 
$\tilde A_{1,2}$
are ${\cal{O}}(\sqrt{l})$ and related by 
\begin{equation}
\tilde{A}_1 = i \tilde{A}_2 ~.
\label{A1A2}
\end{equation}
Notice that although we have the same exponential behavior as in
Sec.~\ref{Sectsu2}, the orders of magnitude are different. This is
because, if the orders of magnitude of $\tilde A_{1,2}$ are bigger
than those in Eq.~(\ref{A'12order}), then by Eq.~(\ref{EqnM'}) they will
cause the variation of $M'_a$ around the boundary to exceed ${\cal
O}(l)$. From Eq.~(\ref{EqnA3'}) we get
\begin{equation}
A'_3={\cal O}(l^{3/2}/D^{1/2}) ~.
\label{A'3order}
\end{equation}
{}From Eq.~(\ref{EqnM'}) and the orders of magnitude in
Eqs.~(\ref{A'12order})
and (\ref{A'3order}), we can see that, for $t<-t_D/2$, the $M'_a$
satisfy the $m$-monopole Nahm equation.\footnote{The factor of $1/2$
arises because of the difference in the orders of magnitude of $\tilde
A_a$ between here and Sec.~\ref{Sectsu2}.}

However, the value of $A_3$ before the unitary transformation of
Eq.~(\ref{utran}) does not have to be as small as in Eq.~(\ref{A'3order}),
but can instead contain a constant term\footnote{The $A_3$ cannot be
bigger than ${\cal O}(\sqrt{lD})$, since otherwise it would cause the
$M_a$ to vary too much near the boundary $t=0$, as we will see from
Eq.~(\ref{Steps}).}  of ${\cal O}(\sqrt{lD})$. This general form of
$T^L_a$ can be obtained by doing the inverse of the unitary
transformation on $T'_a$. This gives
\begin{eqnarray}
A_{1,2} &=&  {\tilde A}_{1,2} \sqrt{2D}
e^{-D|t|} + {\cal O}(l^{3/2}/D^{1/2}) \label{A12j} \\
A_3 &=& {\tilde A}_3 \sqrt{D/2} + {\cal O}(l^{3/2}/D^{1/2}) ~.
\label{A3j}
\end{eqnarray}
The shift caused by this unitary transformation is no longer
negligible, as it was in Secs.~\ref{Sectsu2} and \ref{Sectpoles}.
{}From Eq.~(\ref{Mshift}), it is
\begin{equation}
M_3 = M'_3 + \frac{1}{2} {\tilde A}_3^\dagger {\tilde A}_3 ~.
\label{M3shift}
\end{equation}

{}From Eq.~(\ref{EqnM'}) and the forms of the $A_a$, we can see that
within $(-t_D/2,0)$ the $M_a$ are rapidly varying,
\begin{equation}
M_a(0)-M_a(-t_D/2) = -i\epsilon_{abc}\tilde A_b^\dagger\tilde A_c~.
\label{Steps}
\end{equation}
In the following we will see that for the effective construction
problem the discontinuity
in the Nahm data at the boundary is
\begin{equation}
M_a(0)-M'_a(-t_D/2) = -i\epsilon_{abc} \tilde{A}_b^\dagger \tilde{A}_c +
\frac{1}{2} \tilde{A}_3^\dagger \tilde{A}_3 \delta_{a3} ~,
\label{NahmJump}
\end{equation}
where $M_a(0)=T^R_a(0)$ is the boundary value of the Nahm data $T^R_a$
at $t=0$. Because, as we will show, the rapid variation of $M'_a$ in
the infinitesimally small region 
$(-t_D/2,0)$ has only negligible effects on the construction
solutions in the large $D$ limit, 
$M'_a(-t_D/2)$ is effectively the boundary value of the left side Nahm
data in the reduced problem.

To see this, we need to look at the solutions of the construction
equation. As with the Nahm equations, it is easier to study the
solutions using the
block-diagonalized form of the Nahm data $T'_a$ obtained by the unitary
transformation. We decompose the corresponding construction solution
as 
\begin{equation}
v'=\left 
( \begin{array}{c} w' \\ z' \end{array} 
\right) ~, 
\end{equation}
where $w'$ is a $2m$-dimensional vector and $z'$ is
2-dimensional. The decomposition of Eq.~(\ref{ConsEqn}) is the same as
in Eqs.~(\ref{Eqnw}) and (\ref{Eqnz}):
\begin{eqnarray}
0&=&-\frac{d}{dt}w' + \left[ (-M'_a + r_a) \otimes \sigma_a \right]
w' - ({A'}_a^{\dagger} \otimes \sigma_a)z' ~,
\label{Eqnw2} \\
0&=&-\frac{d}{dt}z' - (A'_a \otimes \sigma_a) w' + \left[ (-b'_a+r_a)
\otimes \sigma_a \right] z' ~.
\label{Eqnz2}
\end{eqnarray}

It is clear from Eqs.~(\ref{A'12order}) and (\ref{A'3order}) that in
the middle of the interval (i.e., $t<-t_D/2$) $w'$ and $z'$ are
decoupled. In the following we will consider two types of
decoupled solutions. We will start in the middle of the interval and
then study their behavior near the boundary.

For the first type, neglecting the terms which vanish as
$D\rightarrow \infty$, we have only the $w'$ components
\begin{equation}
v'=\left( \barr{c} w' \\ 0 \earr \right) ~.
\label{typeone}
\end{equation}
Near the boundary region, the second term in Eq.~(\ref{Eqnz2}) is
${\cal O}(w'~\sqrt{lD})~e^{-D|t|}$. After integrating across the
boundary region, this gives the $z'$ components a contribution of
${\cal O}(w'~\sqrt{l/D})$, which is still negligible compared to
$w'$. Hence, the $w'$ in Eq.~(\ref{typeone}) is a solution of
Eq.~(\ref{Eqnw2}) with the last term ignored. Furthermore, the rapid
${\cal O}(l)~e^{-D|t|}$ variation of the $M'_a$ is restricted to an
interval of width $t_D/2$, which is too small to significantly affect
$w'$.  Hence the $w'$ components in Eq.~(\ref{typeone}) solve the
construction equation defined by the Nahm data ${\hat M_a}$ that are
defined by $\hat M_a=M_a$ for $t<-t_D/2$ and $\hat M_a = {\rm
constant}$ for $-t_D/2<t\le 0$. From this definition, we can see that
at the boundary the $\hat M_a$ are not continuously connected to
$T^R_a$, but instead have a jump given by Eq.~(\ref{NahmJump}).

In the second type of solution, the $w'$ components are much smaller
than the $z'$ components. The latter can be obtained by ignoring the
second term in Eq.~(\ref{Eqnz2}). We have two such solutions, both
similar to those in Eq.~(\ref{type3}). We are interested in the one
that is localized near $t=0$,
\begin{equation}
z' = S \left
( \begin{array}{c} 0 \\ \sqrt{2D}e^{-D|t|} \end{array} \right) ~, 
\label{localz}
\end{equation}
where $S$
is $t$-independent. To get the boundary behavior of the $w'$
component, we plug this into Eq.~(\ref{Eqnw2}) and obtain
\begin{equation}
0=-\frac{d}{dt} w' - (2D e^{-2D|t|}) S {a'}^{\dagger} ~,
\label{localeqn}
\end{equation}
where we assumed the large $D$ limit and defined
\begin{eqnarray}
{a'}^{\dagger} = \left( \barr{c} \tilde A_1^\dagger -i \tilde
A_2^\dagger \\ 0 \earr \right) ~.
\label{a'dagger}
\end{eqnarray}
Thus, this zero-mode is 
\begin{equation}
v'_{\rm jump}=S \left( \begin{array}{c} -{a'}^\dagger e^{-2D|t|} \\0\\ 
\sqrt{2D} e^{-D|t|} 
\end{array} \right)
\label{typetwo}
\end{equation}
In contrast with the localized
solutions studied in the previous cases, the lower two components of
$v'_{\rm jump}$ terminate at $t=0$ and need not satisfy any boundary
conditions. Hence, in this case $v'_{\rm jump}$ gives a linearly
independent solution that is concentrated within an interval of width
$1/D$ adjacent to the boundary. In the $D\rightarrow \infty$ limit, it
is orthogonal to the other solutions and has norm $S^\dagger S$. Note
that while the $w'$ components are of order $\sqrt{l/D}$
smaller than $z'$, they are ${\cal O}(l)$ at the boundary and cannot
be neglected.

To get the general solution of the original problem, we must
do an inverse unitary transformation
\begin{eqnarray}
v' \rightarrow v= \left[ \left( \barr{cc} 1 & A^\dagger_3/D \\ \\
 -A_3/D
& 1 \earr \right) \otimes I_2 \right] v'
\end{eqnarray}
on the solutions of Eqs.~(\ref{typeone}) and (\ref{typetwo}). 

For the first type of solution, Eq.~(\ref{typeone}), $w=w'$ up to
terms that can be neglected.  The $z'$ 
remain small. Hence, in the normalization integral, Eq.~(\ref{su2norm}), and
the field integrals, Eq.~(\ref{su2fields}), the only contributions
come from the $w$ components, which satisfy the construction
equation defined by the $m$-monopole Nahm data $\hat M_a$ defined
below Eq.~(\ref{typeone}).

For the second type of solution, Eq.~(\ref{typetwo}), the $w$
components receive contributions from
the $z$ components after the transformation. We have
\begin{eqnarray}
v_{\rm jump} = S \left( \begin{array}{c} \left( -\tilde A_1^\dagger +
i\tilde A_2^\dagger \right) e^{-2D|t|} \\ \tilde A_3^\dagger e^{-D|t|}
\\0\\  
\sqrt{2D} e^{-D|t|} \end{array} \right) ~.
\label{JumpingMode}
\end{eqnarray}
At the boundary the upper components obey 
$w_{\rm jump}(0)=-Sa^\dagger$, where
\begin{eqnarray}
a^\dagger = \left( \barr{c} \tilde A_1^\dagger -i
\tilde A_2^\dagger \\ - \tilde A_3^\dagger \earr \right)
= \tilde{A}_a^{\dagger} \otimes \sigma_a \left
( \begin{array}{c} 0\\1 \end{array} \right) ~.
\label{JumpingRelation}
\end{eqnarray}
This satisfies
\begin{eqnarray}
-\frac{1}{2}tr_2(a^\dagger a \sigma_a) &=& -\frac{1}{2} \left
( i\epsilon_{abc} \tilde{A}_b^\dagger \tilde{A}_c + \delta_{a3}
\tilde{A}_b^\dagger \tilde{A}_b - \tilde{A}_a^\dagger \tilde{A}_3 -
\tilde{A}_3^\dagger \tilde{A}_a \right) \nonumber \\
&=& -i\epsilon_{abc}
\tilde{A}_b^\dagger \tilde{A}_c + \frac{1}{2} \tilde{A}_3^\dagger
\tilde{A}_3 \delta_{a3} \nonumber \\
&=& T^R_a(0) - \hat M_a(0) ~.
\end{eqnarray}
Equation~(\ref{A1A2}) has been used in the second equality, while
Eq.~(\ref{NahmJump}) and the definition of the $\hat M_a$ have been
used in the third equality.  Comparing with Eq.~(\ref{NahmEqnJump}),
we see that the $a$ defined in Eq.~(\ref{JumpingRelation}) is the
jumping data for the reduced problem.  In the large $D$ limit, the
rapid variation in the $w$ component of Eq.~(\ref{JumpingMode}) over
this interval gives
\begin{equation}
 \Delta w_{\rm jump} = w_{\rm jump}(0)-w_{\rm jump}(-t_D/2) 
     \approx w_{\rm jump}(0) = -S a^\dagger ~.
\label{wjump}
\end{equation}
This is just the discontinuity expected from Eq.~(\ref{ConsEqnJump}).
As noted above, the inner products containing $v_{\rm jump}$ are
dominated by the $z$ components. These produce the second terms on the
right hand sides of Eqs.~(\ref{NormCond}) and (\ref{Fields}).

(In the above discussion, we have mainly concentrated on the region around
the middle boundary $t=0$. While boundary conditions at other
boundaries may eliminate some of the other solutions, they cannot
eliminate $v_{\rm jump}$, because it is localized at $t=0$.)

A general solution of the construction equation is a linear
combination of these two types of solutions.  In the $D\to \infty$
limit, the contribution of the first type to the upper components is
continuous, while that from the second type has a
discontinuity of the form of Eq.~(\ref{wjump}).  Only the second type
of solution has nonzero lower components, and their only effects
are to give contributions to the normalization and field integrals
that are quadratic in $S$.

Thus, we have two different but equivalent ways of looking at
this problem. From the point of view of the original $(m+1,m)$ ADHMN
construction with one monopole far away, the Nahm data $T^L_a$ gives
two types of solutions: those which have negligible lower components,
and the localized solution of Eq.~(\ref{JumpingMode}). From the point
of view of the reduced $(m,m)$ construction, the upper components of
the first type of solution satisfy the construction equation given by the
reduced Nahm data ${\hat M_a}$ and may be discontinuous at the
boundary. The effect of the localized solution is replaced by the
jumping data that describe the discontinuity of the Nahm data ${\hat
M_a}$.

In the reduced problem, the jumping
data is part of the Nahm data. It is interesting to note that,
although the jumping data arise from a localized solution of the
construction equation of the original problem, they are given, through
Eq.~(\ref{JumpingRelation}), by the 
off-diagonal elements $A_a$ of the original Nahm matrices.

\subsection{$k=0$: Disappearance of the jumping data}
\label{Sectk=0}
In this subsection, we start with an $(m,m)$ monopole configuration
that has jumping data. We again consider the limit where one of the
$(1,0)$ monopoles is displaced by a distance $D\gg l$ along the
$z$-axis. Using the same arguments and notation as in
Sec.~\ref{Sectk=1}, we find that the original Nahm data $T^L_a$
generically take the form
\begin{eqnarray}
T^L_{1,2} &=& \left( \barr{cc} {\cal O}(l) & \tilde A_{1,2}^\dagger
\sqrt{2D} e^{-D|t|} + {\cal O}(l^{3/2}/D^{1/2}) \\ \\ \tilde A_{1,2}
\sqrt{2D} e^{-D|t|} + {\cal O}(l^{3/2}/D^{1/2}) & {\cal
O}(l) \earr \right) ~, \nonumber \\ \nonumber \\
T^L_3 &=& \left( \barr{cc} {\cal O}(l) & \tilde A_3^\dagger \sqrt{D/2}
+ {\cal O}(l^{3/2}/D^{1/2})
\\ \\ \tilde A_3 \sqrt{D/2} + {\cal O}(l^{3/2}/D^{1/2}) & D + {\cal
O}(l) \earr \right) ~. \nonumber
\end{eqnarray}
Evaluating this at the boundary $t=0$, we get
\begin{eqnarray}
T^L_{1,2}(t=0)=\left( \begin{array}{cc} {\cal{O}}(l) & \tilde
A_{1,2}^\dagger \sqrt{2D} + {\cal O}(l^{3/2}/D^{1/2}) \\ \\ \tilde
A_{1,2} 
\sqrt{2D} + {\cal O}(l^{3/2}/D^{1/2})& {\cal{O}}(l) \end{array}
\right)~, \nonumber \\ \nonumber \\
T^L_3(t=0)=\left( \begin{array}{cc} {\cal{O}}(l) & \tilde A_3^{\dagger} 
\sqrt{D/2} + {\cal O}(l^{3/2}/D^{1/2}) \\ \\
\tilde A_3 \sqrt{D/2} + {\cal O}(l^{3/2}/D^{1/2})&
D+{\cal{O}}(l)\end{array} \right)~,
\label{T1BD}
\end{eqnarray}
where the $\tilde{A_a}$ are ${\cal{O}}(\sqrt{l})$. We write the
jumping data as
\begin{equation}
a = \left( \barr{cc} \alpha_1 & \beta_1 \\ \alpha_2 & \beta_2 \earr
\right)~, 
\label{aNota}
\end{equation}
where 1,2 are spinor indices, $\alpha_{1,2}$ are $m-1$-dimensional row
vectors and $\beta_{1,2}$ are complex numbers. In this notation, the
discontinuities of the Nahm data, Eq.~(\ref{NahmEqnJump}), are
\begin{eqnarray}
&\Delta& T_1=-\frac{1}{2}\left( \begin{array}{cc} \alpha_1^\dagger
\alpha_2 + \alpha_2^\dagger \alpha_1 & \alpha_1^\dagger \beta_2 +
\alpha_2^\dagger \beta_1 \\ \\ \beta_1^\dagger \alpha_2 +
\beta_2^\dagger 
\alpha_1 & \beta_1^\dagger \beta_2 + \beta_2^\dagger \beta_1
\end{array} 
\right)~, \nonumber \\ \nonumber \\
&\Delta& T_2=\frac{i}{2}\left( \begin{array}{cc} \alpha_1^\dagger
\alpha_2 - \alpha_2^\dagger \alpha_1 & \alpha_1^\dagger \beta_2 -
\alpha_2^\dagger \beta_1 \\ \\ \beta_1^\dagger \alpha_2 -
\beta_2^\dagger 
\alpha_1 & \beta_1^\dagger \beta_2 - \beta_2^\dagger \beta_1
\end{array} 
\right)~,   \label{JumpA} \\ \nonumber \\
&\Delta& T_3=-\frac{1}{2}\left( \begin{array}{cc} \alpha_1^\dagger
\alpha_1 - \alpha_2^\dagger \alpha_2 & \alpha_1^\dagger \beta_1 -
\alpha_2^\dagger \beta_2 \\ \\ \beta_1^\dagger \alpha_1 -
\beta_2^\dagger 
\alpha_2 & \beta_1^\dagger \beta_1 - \beta_2^\dagger \beta_2
\end{array} 
\right)~. \nonumber
\end{eqnarray}
Since all elements of the $T^R_a$ are ${\cal{O}}(l)$ for $t\ge 0$, we
have
\begin{eqnarray}
T^L_a(t=0) + \Delta T_a = {\cal O}(l) ~.
\label{k=0match}
\end{eqnarray}
By
comparing the $mm$-elements of Eqs.~(\ref{T1BD}) and (\ref{JumpA}), we
get
\begin{eqnarray}
\beta_1=\sqrt{2D}+{\cal O}(l/\sqrt{D})~,
~~~~~~~~\beta_2={\cal{O}}(l/\sqrt{D})~, 
\label{aOrder1}
\end{eqnarray}
where an arbitrary phase can be absorbed in the redefinition. 
Using the constraint from Eq.~(\ref{A1A2}), we find from the off-diagonal
blocks in Eqs.~(\ref{T1BD}) and (\ref{JumpA}) that 
\begin{eqnarray}
\alpha_1=\tilde A_3 + {\cal O}(l^{3/2}/D)~,~~~~~~~~
\alpha_2=2\tilde A_1 + {\cal O}(l^{3/2}/D)~.
\label{aOrder2}
\end{eqnarray}
These results imply that the discontinuities in the upper-left
$(m-1)\times (m-1)$
matrices in Eq.~(\ref{JumpA}) can be written as
\begin{eqnarray}
\Delta M_a = i\epsilon_{abc} \tilde A_b ^\dagger \tilde A_c -
\frac{1}{2} \tilde A_3 ^\dagger \tilde A_3 \delta_{a3} + {\cal
O}(l^2/D)~. 
\label{DU}
\end{eqnarray}

As before, by means of a unitary transformation we can make the
$T^L_a$ block diagonal away from the
boundary and then define
$(m-1)\times(m-1)$ Nahm data $\hat M_a$. The effective difference
between the $\hat M_a$ and the $M_a$ at the boundary 
is exactly $-1$ times the quantity in Eq.~(\ref{DU}), as we saw in
Eq.~(\ref{NahmJump}). Thus, by using the
$\hat M_a$ and the $T^R_a$, the jumping data effectively 
disappear in the large $D$ limit.

As in Sec.~\ref{Sectk=1}, all but one of the solutions of the
construction equation have upper components $w$ that solve the
construction equation associated with $\hat M_a$ and lower components
$z$ that are negligible. The remaining solution is localized near
$t=0$ and is of the form
\begin{equation}
N \left( \begin{array}{c} {\cal{O}}\left(\sqrt{l}\right)
e^{-D|t|}  \\ 0
\\ \sqrt{2D} e^{-D|t|} \end{array} \right)~,
\label{vJ2}
\end{equation}
where $N$ is constant and we have only indicated the order of the
magnitude of the first $2m-2$ components. From the jumping data in
Eqs.~(\ref{aOrder1}) and (\ref{aOrder2}), we see that the
discontinuities in the construction equation solutions at $t=0$ must
be of the form
\begin{equation}
S \left( \begin{array}{c} {\cal{O}}\left(\sqrt{l}\right)
\\ \sqrt{2D} \\ 0 \end{array} \right) ~,
\label{discon}
\end{equation}
where the notation is the same as in Eq.~(\ref{vJ2}). 
In order to connect properly to the solutions on
the right interval, both 
$N$ and $S$ have to be proportional to $1/\sqrt{D}$. Hence, neither
the localized solution nor the jumping data contribute to the
normalization or field integrals in the $D\rightarrow \infty$
limit. Furthermore, in this limit the upper $2m-2$ components of the
solutions become continuous at the 
boundary. Thus, the fields become the same as in
the $(m-1,m)$ construction.

\section{An explicit SU(3) example} 
\label{Secteg}
We illustrate the results of Sec.~\ref{Sectpoles} and \ref{Sectjdata}
by an example with SU(3) broken to U(1) $\times$ U(1) with asymptotic
Higgs field $\Phi = {\rm diag}(t_1,t_2,t_3)$.  We consider the $(2,1)$
monopole solution, whose explicit Nahm data are available. This
example has been discussed in detail in \cite{Houghton:1999qu}. Here
we will be interested in the limit where one fundamental $(1,0)$
monopole is removed. We will choose the coordinates so that one
$(1,0)$ monopole is at the origin and the other $(1,0)$ monopole is a
distance $D$ away on the $z$-axis. The $(0,1)$ monopole is at a distance
of order $l$ from the origin.

The Nahm data in the right interval $(t_2,t_3)$ are simply real
numbers that give the coordinates of the $(0,1)$ fundamental
monopole. The boundary condition requires these constants to be equal
to the boundary values of the 11-elements of the $2\times
2$-dimensional Nahm data defined in the left interval $(t_1,t_2)$.  If
all three monopoles are collinear,\footnote{If they are not collinear,
we must do a unitary transformation on $T_a^L$, as we will see
later in this section.} then up to spatial rotations and translations the
Nahm data in $(t_1,t_2)$ are \cite{Dancer:kn}
\begin{equation}
T^L_a=\frac{1}{2} f_a \sigma_a + \frac{D}{2} \delta_{a3} \qquad $(no
sum)$
\label{DancerT}
\end{equation}
for $a=1,2,3$, where the $\sigma_a$ are Pauli matrices. The analytic
functions $f_a$ are
\begin{eqnarray}
f_1(t)&=& -\frac{D cn_k(D(t-t_1))}{sn_k(D(t-t_1))}~, 
\nonumber  \\ \nonumber \\
f_2(t)&=& -\frac{D dn_k(D(t-t_1))}{sn_k(D(t-t_1))}~,\label{DancerF} 
 \\ \nonumber \\
f_3(t)&=& -\frac{D}{sn_k(D(t-t_1))}~,  \nonumber 
\end{eqnarray}
where $0 \le k \le 1$ and $D < \frac{2}{t_2-t_1}K(k)$. 
Here $sn_k$, $cn_k$ and $dn_k$ are the Jacobi elliptic functions and
$K(k)$ is the complete elliptic integral of the first kind.
The $f_a$ have poles at $t_1$ and $t_* = t_1 +
2K(k)/D > t_2$.

In general, the relations between the parameters $k$ and $D$ and the
physical quantities are complicated. They become simple when $k
\rightarrow 1$ in the large $D$ limit. 
Within the pole region
$t-t_1 \ll 1/D$, we have
\begin{eqnarray}
T^L_a = \frac{\sigma_a}{2(t-t_1)} + {\cal O}(D^2(t-t_1)) ~.
\label{Tappr1}
\end{eqnarray}
Away from this region, $t-t_1 \gg 1/D$, we can use the approximation
\begin{equation}
f_1=f_2 = -D / \sinh (D(t-t_1)) ~,~~~~ f_3=-D \coth (D(t-t_1)) ~,
\label{k=1lim}
\end{equation}
valid when $k\rightarrow 1,$ to obtain
\begin{eqnarray}
T^L_1 &=& \left( \barr{cc} 0&-De^{-D|t-t_1|}\\ \\-De^{-D|t-t_1|}&0 \earr
\right)  ~, \nonumber \\ \nonumber \\
T^L_2 &=& \left( \barr{cc} 0& i De^{-D|t-t_1|}\\ \\ -i De^{-D|t-t_1|}&0
\earr 
\right)  ~, \label{Tappr2} \\ \nonumber \\
T^L_3 &=& \left( \barr{cc} -De^{-2D|t-t_1|}&0\\ \\0&D+De^{-2D|t-t_1|}
\earr \right) ~, \nonumber 
\end{eqnarray}
for $t < (t_1+t_*)/2$. (Higher order exponential terms have been
omitted here, and in later similar expressions.) This explicitly
verifies the predictions of Eqs.~(\ref{hAform2})-(\ref{Aorder2}). 

The expressions in the region $t > (t_1+t_*)/2$ can be
related to those in $t < (t_1+t_*)/2$ by
\begin{eqnarray}
sn_k(x) &=& sn_k(2K(k)-x) ~,  \nonumber \\
dn_k(x) &=& dn_k(2K(k)-x) ~,  \\
cn_k(x) &=& -cn_k(2K(k)-x) ~. \nonumber
\end{eqnarray}
In particular, the upper-left element of $T^L_3$ has to be ${\cal
O}(l)$ at $t=t_2$. Then we have
\begin{eqnarray}
T^L_1 =& \left( \barr{cc} 0&De^{-D|t-t_*|}\\ \\De^{-D|t-t_*|}&0 \earr 
\right) &= \left( \barr{cc} 0&\sqrt{Dc}e^{-D|t-t_2|} \\ \sqrt{Dc}
e^{-D|t-t_2|}&0 \earr \right)  ~,\nonumber \\ \nonumber  \\ 
T^L_2 =& \left( \barr{cc} 0& i De^{-D|t-t_*|}\\ \\-i De^{-D|t-t_*|}&0
\earr \right) 
&= \left( \barr{cc} 0& i\sqrt{Dc}e^{-D|t-t_2|} \\ \\-i\sqrt{Dc}
e^{-D|t-t_2|}&0 \earr \right)  ~,\label{Tappr3}\\ \nonumber \\
T^L_3 =& \left( \barr{cc} -De^{-2D|t-t_*|}&0\\ \\0&D+De^{-2D|t-t_*|}
\earr 
\right) &= \left( \barr{cc} -ce^{-2D|t-t_2|}&0\\ \\0& D+ce^{-2D|t-t_2|}
\earr \right) ~, \nonumber
\end{eqnarray}
for $t > (t_1+t_*)/2$, where $c=De^{-2D|t_*-t_2|}$ is a positive
constant of
${\cal O}(l)$ that gives the position of the $(0,1)$ monopole at
$(0,0,-c)$. These 
are examples of Eqs.~(\ref{A'12order}) and
(\ref{A'3order}). 

Away from the boundaries, the Nahm data is
approximately diagonal, with
\begin{equation}
T^L_1=T^L_2=0, ~~~ T^L_3=\left
( \begin{array}{cc} 0&0\\0&D \end{array} \right) ~.
\end{equation}
This corresponds to
two widely separated $(1,0)$ monopoles, with one at the origin and the
other at $D$ on the $z$-axis.

Eq.~(\ref{Eqnz2}) gives two solutions localized near a boundary, with
lower components 
\begin{equation}
z_-=\sqrt{2D} \left
( \begin{array}{c} 1\\0 \end{array} \right) e^{-D|t-t_1|} ~,
~~~~~~
z_+=\sqrt{2D} \left( \begin{array}{c} 0\\1 \end{array} \right)
e^{-D|t-t_2|} ~.
\end{equation}
The second one is localized near $t_2$. Substituting 
this into Eq.~(\ref{Eqnw2}) gives 
\begin{equation}
w = \left
( \begin{array}{c} 
-\sqrt{2c} e^{-2D|t-t_2|}\\0 \end{array} \right) ~.
\label{su3w}
\end{equation}
Although this has a
nonzero value only at the boundary in the $D\rightarrow \infty$
limit, it provides an extra degree of freedom for
connecting the solutions from the two intervals. Evaluating
Eq.~(\ref{su3w}) at the boundary 
$t_2$ gives the jumping data $a^\dagger$, with the correct jump
$\Delta T_a = -\frac{1}{2}tr_2(a^\dagger 
a \sigma_a)=-c \delta_{a3}$. 

So far we have restricted ourselves to the case where all three
monopoles are collinear.  We now relax this restriction.  Without loss
of generality, we can rotate the system so that the $(1,0)$ monopoles
are on the $z$-axis while the $(0,1)$ monopole is in the $xz$ plane at
$(x,0,z)$.  To obtain the Nahm data for this solution, we perform a
unitary transformation using
\begin{equation}
{\cal{U}} =
\left( \begin{array}{cc} \cos{\frac{\theta}{2}} &
\sin{\frac{\theta}{2}} \\ \\ -\sin{\frac{\theta}{2}} &
\cos{\frac{\theta}{2}} \end{array} \right)
\end{equation}
on the previous Nahm data, with
\begin{eqnarray}
\sin\theta=x/\sqrt{Dc}~,~~~{\rm
where}~~c=(\sqrt{z^2+x^2}-z)/2 \equiv (r-z)/2 ~.
\end{eqnarray}
This unitary transformation
rotates the 11-elements at $t_2$ (i.e.~the position of the $(0,1)$
monopole) along an ellipsoid with foci $(0,0,0)$ and $(0,0,D)$ on the
$z$-axis \cite{Lee:1997ny}. The Nahm data
after the transformation are (in the $D\rightarrow \infty$ limit)
\begin{eqnarray}
T^L_1 &=& \left( \begin{array}{cc} xe^{-D|t-t_2|} &
\high{ \sqrt{\frac{D(r-z)}{2}}~e^{-D|t-t_2|} }  \\ \\
\high{ \sqrt{\frac{D(r-z)}{2}}~e^{-D|t-t_2|} } & -xe^{-D|t-t_2|}
\end{array} \right)~, \nonumber \\ \nonumber\\ 
T^L_2 &=& \left( \begin{array}{cc} 0 &
\high{ i\sqrt{\frac{D(r-z)}{2}}~e^{-D|t-t_2|} }\\ \\
\high{ -i\sqrt{\frac{D(r-z)}{2}}~e^{-D|t-t_2|} } & 0 \end{array}
\right)~, \\ \nonumber \\
T^L_3 &=& \left( \begin{array}{cc} \high{ \frac{x^2}{2(r-z)} -
\frac{r-z}{2}~e^{-2D|t-t_2|} } & 
\high{ \sqrt{\frac{D}{2}} \frac{x}{\sqrt{r-z}} } \\ \\ \high
{ \sqrt{\frac{D}{2}} \frac{x}{\sqrt{r-z}} } & 
\high{ D-\frac{x^2}{2(r-z)} + \frac{r-z}{2}~e^{-2D|t-t_2|} }
\end{array} \right)
~. \nonumber
\end{eqnarray}
These are examples of Eqs.~(\ref{A12j}) and (\ref{A3j}). 
This transformation changes Eq.~(\ref{su3w}) to
\begin{equation}
\left( \begin{array}{c} -\sqrt{r-z}e^{-2D|t-t_2|} \\
\high{ \frac{x}{\sqrt{r-z}} e^{-D|t-t_2|} } \end{array} \right) ~,
\end{equation}
again giving the correct
jumps $\Delta T_1 = x, \Delta T_2=0, \Delta T_3= z$.

\section{The D-brane picture}
\label{Sectdpic}
As mentioned in the introduction, in the D-brane picture the monopoles
are D-strings stretching between adjacent D3-branes. From the
perspective of the D3-brane, the endpoints of the D-strings are
magnetic sources that generate magnetic flux in the three spatial
directions on the D3-brane.

We denote the spatial distance on the D3-brane as $r$ and the Higgs
field as $\Phi$. The Higgs field of the D3-brane describes its
transverse fluctuations (which we denote as $\hat t$ ) up to a factor
of $2\pi l_s^2$ (where $l_s$ is the string scale).  The nontrivial
profiles of the Higgs fields of the monopoles describe the bending of
the D3-brane due to the presence of the D-strings
\cite{Callan:1997kz,Gibbons:1997xz}. If we consider the case where the
D-strings end at the D3-brane, one typically finds that spikes are
created at the D-string endpoints.  These spikes can be obtained from
the asymptotic Higgs profile of the charge $k$ solution
\begin{equation}
|\hat t|=2\pi \l_s^2 |t|=2\pi \l_s^2 \Phi(r) = \pi \l_s^2
\frac{|k|}{er} ~.
\label{spike}
\end{equation}
(For this section, we have restored the factors of $e$, with
$r_a\rightarrow er_a$, $T_a\rightarrow e T_a$.)  This formula is valid
if $r$ is bigger than the monopole separation scale $l$ and the
monopole core size.

On the other hand, there is a dual description of the above
phenomena from the perspective of the D-strings
\cite{Constable:1999ac}. As we have seen in the ADHMN construction,
the $k \neq 0$ case corresponds to situations where the Nahm data on
the two sides of the boundary have different dimensions. In this case
(for $|k|>1$), we have poles emerging at the boundary. Since the Nahm
data give the transverse fluctuations of the D-strings, these poles,
with $k$-dimensional irreducible SU(2)-valued residues, means that
these D-strings are no longer distinct from each other near the
D3-branes. They form a noncommutative two-sphere \cite{Kabat:1997im}
and have an overall funnel-like geometry.  The
radius of this two-sphere is naturally defined as 
\begin{equation}
R(\hat t)^2 = \frac{4\pi^2 l_s^4}{|k|} \sum_{i=1}^3 {\rm Tr} \left
[ T^i(\hat t)^2 \right] ~.
\label{ncsr}  
\end{equation}
In the pole region $|\hat t|<2\pi l_s^2/el$ we have 
\begin{equation}
R(\hat t)=\pi l_s^2 \frac{|k|}{e |\hat t|} \sqrt{1-\frac{1}{k^2}} ~,
\label{funnel}
\end{equation}
where $\sum_i (J^i)^2 = \frac{1}{4} (k^2-1) I_k$ has been used.

Noticing that the three transverse directions of the D-strings are the
same as the three spatial directions on the D3-brane, we can then
identify $R$ in Eq.~(\ref{funnel}) with $r$ in
Eq.~(\ref{spike}). These two dual descriptions, Eq.~(\ref{spike}) and
Eq.~(\ref{funnel}), of the brane junction agree well for large
$k$. This can be explained as follows \cite{Constable:1999ac}. This
junction can be described by the non-Abelian world-volume Born-Infeld
actions of the D3-branes and D-strings. This will give, respectively,
the BPS monopole equation on the D3-branes and the Nahm equation on
the D-string. The regions of validity of these two descriptions are
restricted to the region where the effect of a derivative on the
fields is less than a factor of $1/l_s$, so that the higher order
string corrections to the Born-Infeld actions can be ignored. This
means $r\gg l_s$ in Eq.~(\ref{spike}) and $|\hat t| \gg l_s$ in
Eq.~(\ref{funnel}). These regions overlap when $|k|\gg1$. In terms of
$\hat t$, the overlapping region is $l_s \ll |\hat t| \ll k\pi l_s/e$.
This overlapping region extends into the pole region of the Nahm
data if $e<2\pi l_s/l$. This is consistent with the weak electric
coupling limit where the monopole description makes sense.

\begin{figure}[tbh]
\begin{center}
\epsfig{file=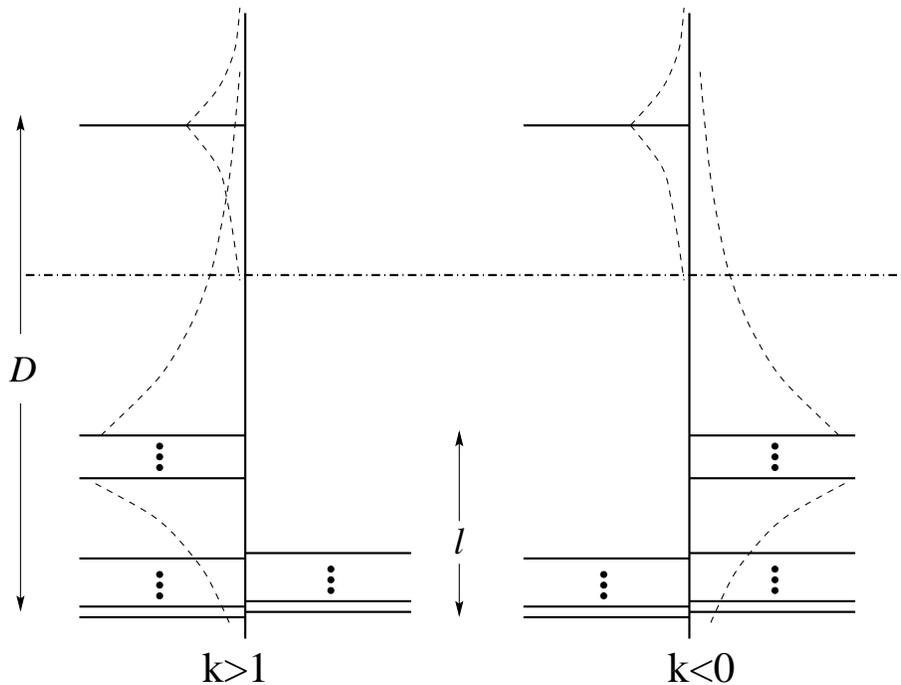, width=12cm}
\end{center} 
\medskip
\caption{The D-brane interpretation of the $k>1$ and $k<0$ cases. In
this and 
the next two figures the perpendicular solid lines represent the
D3-branes and the horizontal solid lines are the D-strings. The
dashed-dotted lines separate the region of size $l$ from that of size
$D$. In this figure the dashed lines represent
the D3-brane spikes.} 
\label{kpolesfig}
\end{figure}

Now the transitions of the poles in Sec.~\ref{Sectpoles} can be
interpreted in this D-brane picture.  For the $k>1$ case, we only see
a $k-1$ net magnetic charge in the spatial region $l\ll r \ll D$ on
the D3-brane. In the dual description on the D-strings, this
corresponds to the $(k-1)$-dimensional poles in the boundary region
$|t| \ll 1/el$. If we move far away, to the spatial region $r \gg D$,
we start to see the charge of the removed monopole. In the dual
description, this corresponds to the higher dimensional poles within
the boundary region $|t| \ll 1/eD$.  The $k<0$ case is similar, except
now the removed monopole cancels one unit of magnetic charge. These
two cases are sketched in Fig.~\ref{kpolesfig}.

We also notice that in these two cases the distant D-string segment
interacts predominantly with the noncommutative two-sphere part of the
D-strings that contribute to the net magnetic charge on the
D3-brane. These string excitations are described by the poles in the
Nahm data. The interactions between this distant D-string segment and
the other D-strings are small. For the $k>1$ case, these are described
by the $F_a$ in Appendix \ref{Appk>1}. These fall as a power of $t$
and vanish at the boundary $t=0$. For the $k<0$ case, the interactions
are described by the $A_a$ in Sec.~\ref{Sectk<0}, which are at most of
order $l$ at the boundary $t=0$.

The ADHMN construction with
jumping data can be obtained by T-duality of the D0-D4 system. One
finds that the jumping data describe excitations of bosonic strings
that stretch between the 
D-strings and D3-branes. These excitations are always localized on
the D3-branes. This method naturally imposes the restriction that the
number of D-string segments should be the same on both sides of
each D3-brane. In other words, from the D3-brane point of view, all
the magnetic flux coming from the D-strings on one side has to exit
within a finite distance to the D-strings on the other side.
We have seen that different boundary conditions can be naturally
linked to the others by removing certain D-string segments. 
Therefore we can start with this configuration and derive
all the other boundary conditions by removing certain
D-string segments. It is
then interesting to see how the appearance or the disappearance of the
jumping data can be interpreted in this D-brane picture. 

In addition, because of the T-duality, the system is compactified
along the D1 direction. 
If we remove away all the D-string segments in one
interval, we effectively decompactify the system.

\begin{figure}[t]
\begin{center}
\epsfig{file=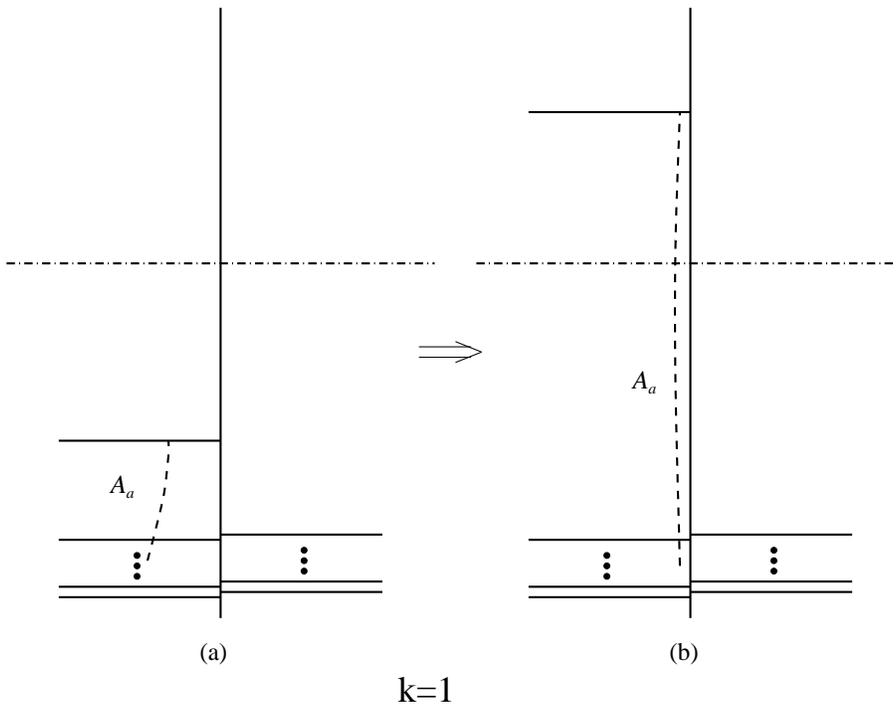, width=12cm}
\end{center} 
\medskip
\caption{The D-brane interpretation of the $k=1$ case. In this figure
the dashed lines represent the fundamental string
excitations between the D-strings; (a) and (b) represent the D-brane
pictures before and 
after removing a D-string segment to a distance $D$, respectively.} 
\label{ke1fig}
\end{figure}

\begin{figure}[t]
\begin{center}
\epsfig{file=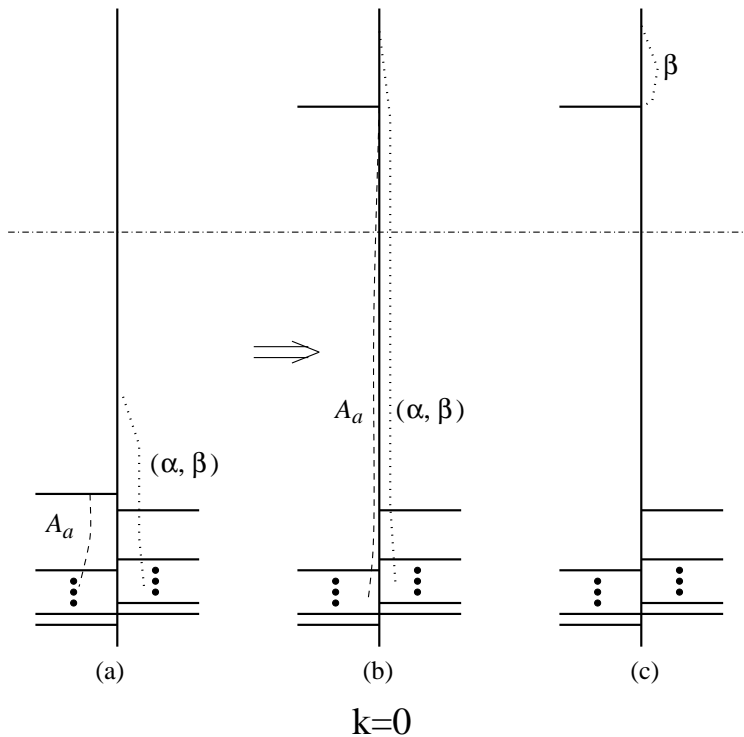, width=10cm}
\end{center} 
\medskip
\caption{The D-brane picture of the $k=0$ case. In this figure
the dashed lines represent the fundamental string
excitations between the D-strings; the dotted line is the
hypermultiplet (jumping data). (a) represents the D-brane
picture before we removing a D-string segment to a distance $D$. We
use (b) and (c) to represent the D-brane picture after this
separation, where (c) is the final effective picture.}
\label{ke0fig}
\end{figure}

We first look at the $k=1$ case. We focus on the fields describing the
fundamental strings stretching between the D-string segment we want to
separate and the rest of the D-strings. In the Nahm data, these fields
are the $A_a$ of Sec.~\ref{Sectk=1}. Before the separation, they are
within a region of size $1/el$ (in terms of the D-string coordinate
$t$) adjacent to the D3-brane. As we remove the D-string segment to a
distance of order $D$ (in terms of the distance on the D3-brane),
these fields become localized closer to the D3-brane; they are
effectively restricted to a region of order $1/eD$ as indicated by the
$e^{-D|t|}$ dependence of the $A_a$ in Eqs.~(\ref{A12j}) and
(\ref{A3j}). Therefore, as $D \rightarrow \infty$ these string
excitations will be restricted to the D3 brane and effectively appear
as interactions between the D-strings and the D3-brane, just like the
hypermultiplet. Quantitatively, as we can see from
Eq.~(\ref{JumpingRelation}), the effective jumping data is indeed
proportional to the $A_a$, with a normalization factor of order
$\sqrt{D}$. Notice that, in contrast with the previous two cases,
these string excitations are not small, even though the system left is
neutral. They are of order $\sqrt{lD}$ at $t=0$. This case is sketched
in Fig.~\ref{ke1fig}.

The $k=0$ case is a bit more complicated.  Just as in the above $k=1$
case, as we remove one D-string segment the fundamental string modes
connecting this D-string segment and the other D-strings become
restricted to the D3 brane (see (b) in Fig.~\ref{ke0fig}). One the
other hand, we also have the jumping data between the D-string and D3
branes (see (b) in Fig.~\ref{ke0fig}). Equation~(\ref{k=0match})
requires that the leading, ${\cal O}(\sqrt{lD})$, parts of these two
contributions cancel. This leaves the $\beta$ part of the jumping
data, represented by (c) in Fig.~\ref{ke0fig}. This decouples in the
$D\rightarrow \infty$ limit.

\section{Conclusions and discussion}
\label{SectConclu}
In this paper, we have related the boundary conditions in the ADHMN
constructions for different magnetic charges by removing fundamental
monopoles one by one to spatial infinity; we were particularly
interested in the cases involving jumping data, which seem quite
different from the other cases. We demonstrated the equivalence in the
$D\rightarrow \infty$ limit between the ADHMN construction for the
original $(\dots,m+k,m,\dots)$ problem, with one $(\dots,1,0,\dots)$
fundamental monopole removed a distance $D$ from the others, and the
reduced $(\dots,m+k-1,m,\dots)$ problem.

For the Nahm data $T^L_a$ on the left interval, we generally find a
small region near the boundary whose width $\epsilon_D$ goes to zero
as $D \rightarrow \infty$. Away from this boundary region, we can make
the $T^L_a$ block-diagonal with $D\delta_{a3}$ in one block and the
decoupled Nahm data of the reduced problem in the other. The values of
the $T^L_a$ at $t=-\epsilon_D$ become the boundary values for the
reduced problem. The Nahm data generally vary rapidly in $-\epsilon_D
< t <0$. This can give changes in the pole behavior or can lead to an
effective discontinuity $\Delta T_a$.

For $k>1$, the dimensions of the pole terms change from $k-1$ to $k$
in the boundary region; this variation does not affect the fields of
the reduced problem.  For $k=1$, this rapid variation gives an
effective discontinuity between the Nahm data on the two sides of the
boundary in the large $D$ limit. The construction equation for this
case has a solution localized in this small region that has the same
effect as the jumping data in the reduced problem. For $k=0$, the
rapid variation of $T^L_a$ cancels the effect of the original jumping
data and makes the reduced Nahm data continuous across the boundary
between the left and right intervals.  For the $k<0$ case, removing
a monopole on the left also causes changes on the right interval:
The $|k|$-dimensional pole terms of the original $T^R_a$ are
restricted to $0<t<\epsilon_D$ and go over to the
$(|k|+1)$-dimensional pole terms of the reduced problem when
$t>\epsilon_D$. As in the $k>1$ case, this small region has no effect
on the fields.

In terms of the D-brane picture, removing massive fundamental
monopoles corresponds to removing D-string segments. The transition
between different types of boundary conditions can then be interpreted
in terms of the interactions between the distant D-string segment and
the rest of the system.

An interesting extension of this work would be to consider the case
where some D3-branes coincide with each other. In terms of the
world-volume theory on the D3-branes, this corresponds to having a
non-Abelian unbroken gauge symmetry. In these cases one finds
solutions with clouds of non-Abelian fields surrounding one or more
massive monopoles. These clouds can be interpreted as massless
monopoles with non-Abelian magnetic charge
\cite{Dancer:kn,Lee:1997ny,Weinberg:jh,Lee:1996vz,Weinberg:1998hn,Houghton:2002bz}.

\begin{figure}[htb]
\begin{center}
\epsfig{file=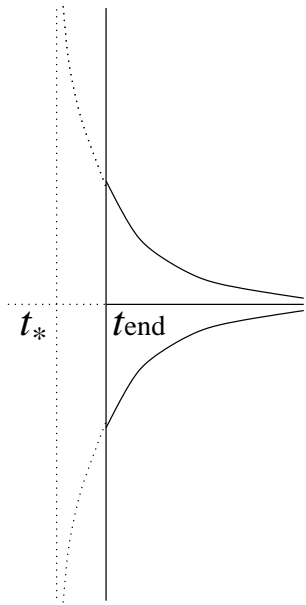, width=4cm}
\end{center} 
\medskip
\caption{The pole behavior for the non-Abelian cloud case. The dotted
lines are extensions of the solid lines to the pole.}
\label{cloudpole}
\end{figure}

In terms of the Nahm data, these clouds have been found to arise in
two different ways. In the first, the clouds originate from the pole
behavior of the Nahm data. An example is the $([1],2)$
solution\footnote{The square bracket denotes a massless monopole.} in
SU(3) $\rightarrow$ SU(2) $\times$ U(1).  The D-brane picture is
sketched in Fig.~\ref{cloudpole}, where $t_{end}=t_1=t_2$ is the
position of the two coinciding D3-branes. The pole of the Nahm data is
at $t_*<t_{end}$. Sending the massless monopole far away corresponds
to bringing $t_*$ closer to $t_{end}$. The size of the cloud, which
can be interpreted as the distance to the massless monopole, is
approximately \cite{Houghton:2002bz}
\begin{equation}
a \approx \frac{1}{2e(t_{end}-t_*)} ~.
\label{cloudsize}
\end{equation}
We can understand this result heuristically by applying the large-$k$
D-brane analysis of Sec.~\ref{Sectdpic} to the present $k=2$ case.  On
the D3-branes, the non-Abelian Higgs fields of the known BPS solution
\cite{Dancer:kn,Weinberg:jh} give $t-t_{end} \approx
\high{\frac{a}{2e(r+a)r}}$ for $a$ much greater than the massive
monopole separation. For $a\gg r$ this gives
\begin{eqnarray}
r \approx \frac{1}{2e(t-t_{end})+1/a} ~.
\end{eqnarray}
On the D-strings, since the pole position is at $t_*$, the D-string
fluctuation scale
\begin{eqnarray}
R \approx \frac{1}{2e} \frac{1}{(t-t_{end})+(t_{end}-t_*)} ~.
\end{eqnarray}
Setting $R=r$ then gives Eq.~(\ref{cloudsize}), up to a factor of
${\cal O}(1)$.

In the second case, the clouds arise from the jumping data. An example
is the $(1,[1],1)$ solution in SU(4) $\rightarrow$ U(1) $\times$ SU(2)
$\times$ U(1). The relation between the cloud and the D-brane picture
is less clear for this situation, as well as for the more complicated
examples where both jumping data and pole behavior contribute to the
clouds \cite{Houghton:2002bz}.

In this non-Abelian case the process of taking individual monopoles to
spatial infinity is more complicated than when the symmetry is
maximally broken.  Because the long-range interactions between
monopoles with non-Abelian charges are more complex than when only
Abelian magnetic charges are present, it is not always possible to
deform a solution into one that is essentially a superposition of
widely separated component monopoles.  A striking example of this is
seen when one examines solutions with massive and massless component
monopoles whose overall magnetic charge is purely Abelian.  Previous
studies \cite{Dancer:kn,Lee:1997ny,Weinberg:1998hn,Houghton:2002bz} of
such ``magnetically color-neutral'' solutions have found that the
massive monopoles must always be enclosed by one or more clouds.  As a
result, one cannot remove one of the massive monopoles to spatial
infinity without simultaneously making one of the clouds infinite in
size.

However, this is not necessarily the case for solutions that are not
magnetically color-neutral.  An example of this is provided by the
solutions with charge $([1],m)$ for SU(3) broken to SU(2)$\times$U(1).
{}From the form of the Nahm construction for these solutions, it
is easy to see that we can use the methods of this paper to remove
$m-2$ of the $([0],1)$ massive fundamental monopoles one by one,
while still maintaining a finite size cloud.  It is only when
we are left with the color-neutral $([1],2)$ solution that this
procedure breaks down.

This process of removing massive monopoles can help us understand
the parameters entering these solutions.  Both the counting and the
interpretation of these are more complicated than when all the
monopoles are massive.  With maximal symmetry breaking, the moduli
space of $(1,m)$ solutions is $4(m+1)$-dimensional, with the
parameters corresponding to three position and one U(1) collective
coordinate for each of the massive monopoles.

With the non-maximal breaking, this counting still holds for the $m=2$
case, but the meaning of some of the parameters is changed.  Eight are
the usual position and U(1) parameters for the two massive monopoles,
but three others correspond to global SU(2) rotations, while the last
characterizes the size of the non-Abelian cloud.  For $m \ne2$, the
dimensions of the spaces have been found to be 6 for $m=1$ and $4m+6$
for $m>2$ \cite{Murray:zk}. (There is no solution for $m=0$.)  By
examination of the explicit $([1],1)$ solutions [which are gauge
equivalent to the $([0],1)$ ones] one sees that four parameters are
position and U(1) variables, while the remaining two dimensions are
due to global SU(2) rotations that correspond to non-normalizable zero
modes.  To understand the parameters when $m>2$, we apply the methods
of this paper to deform the generic solution into one containing $m-2$
massive monopoles that are well separated from each other and from a
$([1],2)$ color-neutral configuration.  It then seems clear that the
parameters should be understood as $4(m-2)$ position and U(1)
variables for the isolated massive monopoles, the twelve described above
for the $([1],2)$ configuration, and two more corresponding to
non-normalizable global SU(2) modes.

The generalization of this procedure to other charges and groups, as
well as to solutions containing more than one cloud, remains a challenge
for future work.

\acknowledgments 
This work was supported in part by the U.S. Department of Energy.

\appendix
\section{}
\label{Appk>1} 
In this appendix we describe the details of the analysis of the Nahm
data and solutions of the construction
equation for the $k>1$ case described in Sec.~\ref{Sectk>1}. We will
study the case of charge
$(m+k,m)$ with one $(1,0)$ monopole removed by a distance $D$, and
compare it with the case of charge $(m+k-1,m)$. We will focus on the
behavior in the neighborhood of the middle boundary at $t_2=0$,
ignoring the boundary regions near $t_1$ and $t_3$.

We decompose
the Nahm data $T^L_a$ and construction equation solution $v$ as
\begin{equation}
T^L_a = \left( \begin{array}{ccc} N_a & E_a^\dagger & F_a^\dagger \\ 
E_a & P_a & G_a^\dagger \\ F_a & G_a & b_a \end{array} \right) ~,
~~~~~
v = \left( \begin{array}{c} u \\ w \\ z \end{array} \right) ~,
\label{k>1form}
\end{equation}
where in each case the first entry is $m$-dimensional, the second is
$(k-1)$-dimensional and the third is one-dimensional. 

The Nahm equations separate into
\begin{eqnarray}
\frac{dN_a}{dt} &=& -i\epsilon_{abc} \left( N_b N_c + E_b^\dagger E_c +
F_b^\dagger F_c^\dagger \right) ~,
\label{EqnNa} \\
\frac{dE_a}{dt} &=& -i\epsilon_{abc} \left( E_b N_c + P_b E_c +
G_b^\dagger F_c \right) ~,
\label{EqnEa} \\
\frac{dF_a}{dt} &=& -i\epsilon_{abc} \left( F_b N_c + G_b E_c +
b_b F_c \right) ~,
\label{EqnFa} \\
\frac{dP_a}{dt} &=& -i\epsilon_{abc} \left( E_b E_c^\dagger + P_b P_c +
G_b^\dagger G_c \right) ~,
\label{EqnPa} \\
\frac{dG_a}{dt} &=& -i\epsilon_{abc} \left( F_b E_c^\dagger + G_b P_c +
b_b G_c \right) ~,
\label{EqnGa} \\
\frac{db_a}{dt} &=& -i\epsilon_{abc} \left( F_b F_c^\dagger + G_b
G_c^\dagger \right) ~.
\label{Eqnba}
\end{eqnarray}

{}From Eq.~(\ref{MatchingCond}) we see that
in the interval $(-1/D,0)$ the matrix 
\begin{equation}
\left( \begin{array}{cc} P_a & G_a^\dagger \\ G_a & b_a \end{array}
\right) \approx -\frac{J_a^{(k)}}{t} ~,
\end{equation}
where the $J_a^{(k)}$ are a $k$-dimensional irreducible representation
of SU(2). The behavior of the other elements of the $T^L_a$ in this
region can also be seen from Eq.~(\ref{MatchingCond}). For $t\le -1/l$, we
have $b_3= D$ and the other elements ${\cal
O}(l)$, just as in Sec.~\ref{Sectsu2}. 

By a unitary transformation using
\begin{equation}
U=\left( \begin{array}{ccc} I_{m\times m}+{\cal O}(l^2/D^2) & {\cal
O}(l^2/D^2) & 
-{\cal F}_3^\dagger/D \\ \\ {\cal O}(l^2/D^2) & I_{(k-1)\times(k-1)}
+{\cal O}(l^2/D^2) & 
-{\cal G}_3^\dagger/D \\ \\ {\cal F}_3/D & {\cal G}_3/D & 1+{\cal
O}(l^2/D^2) \end{array} 
\right) 
\end{equation}
we can subtract constants ${\cal F}_3$ and ${\cal G}_3$ from $F_3$ and
$G_3$, respectively. We choose these constants so that $F_3$ and
$G_3$ are exponentially small in the middle of the interval. Since, as
we will see, ${\cal F}_3$ and ${\cal G}_3$ are both ${\cal O}(l)$, the
effects of the transformation on the other elements of the $T^L_a$ are
negligible. 

Using arguments similar to those applied to the $A_a$ in
Sec.~\ref{Sectsu2}, we can show that for $t<-1/D$ the first and second
components of Eq.~(\ref{EqnGa}) are dominated by the $b_3 G_c$
term. This gives the $t$-dependence of the $G_{1,2}$ to be ${\cal
O}(D) e^{-D|t|}$, where the coefficient must be ${\cal O}(D)$ so that
it can match on to the pole at $t=-1/D$. This, together with the third
component of Eq.~(\ref{EqnGa}), gives $G_3 \sim {\cal O}(l)
e^{-D|t|}$~.  The first and second components of Eq.~(\ref{EqnFa}) are
dominated by the $G_{1,2} E_c$ or $b_3 F_c$ terms. These give $F_{1,2}
\sim {\cal O}(l) e^{-D|t|}$. The third component of Eq.~(\ref{EqnFa})
then implies $F_3 \sim {\cal O}(l) e^{-D|t|}$.

The orders of magnitude of $G_{1,2}$ justify our previous statement
that ${\cal F}_3$ and ${\cal G}_3$ are at most ${\cal O}(l)$, because
otherwise Eq.~(\ref{EqnEa}) or Eq.~(\ref{Eqnba}) would imply that 
$E_{1,2}$ or $b_{1,2}$ would be too big.

For $t<-t_D$, the $F_a$ and $G_a$ are exponentially small and can be
ignored in Eqs.~(\ref{EqnNa}), (\ref{EqnEa}), and (\ref{EqnPa}). These
three are the 
Nahm equations for the $(m+k-1)$-monopole case. The poles 
of the matrices $P_a$ in $(-1/D,0)$ extend into this region and
dominate the right-hand sides of Eqs.~(\ref{EqnEa}) and
(\ref{EqnPa}). This gives the $P_a$ poles whose residue is a
$(k-1)$-dimensional irreducible SU(2) representation in
$(-1/l,-t_D)$. As in Eq.~(\ref{MatchingCond}), we have $E_a
\sim t^{(k-2)/2}$. 
Finally Eq.~(\ref{Eqnba}) can be used to refine the behavior of
$b_a$ using the above information. We find that $b_{1,2} = {\cal
O}(l) + {\cal
O}(l)e^{-2D|t|}$ and $b_3 = D+{\cal O}(D)e^{-2D|t|}$. 

These results are summarized in Table \ref{k>1table}.

\vskip 1cm 
\begin{center}
\begin{tabular}{clcccccccr}
\hline
  & $t_1$ &  & $-1/l$ &  & $-t_D$ &  & $-1/D$ &  & 0
\\ \hline
$N_a$~~ & $\mid\leftarrow$ & & &  &
   ${\cal O}(l)$ & & & & $\rightarrow\mid$
\\
$E_a$~~ & $\mid\leftarrow$ & ${\cal O}(l)$ &
   $\rightarrow\mid\leftarrow$ & ${\cal O}(t^{(k-2)/2})$ &
  $\rightarrow\mid\leftarrow$ & transition & 
$\rightarrow\mid\leftarrow$ & 
${\cal O}(t^{(k-1)/2})$ & $\rightarrow\mid$ 
\\
$F_a$~~ &  $\mid\leftarrow$ & & & ${\cal O}(l) e^{-D|t|}$ &
  $\rightarrow\mid\leftarrow$ & transition &
$\rightarrow\mid\leftarrow$ & 
${\cal O}(t^{(k-1)/2})$ & $\rightarrow\mid$  
\\
$P_a$~~ & $\mid\leftarrow$ & ${\cal O}(l)$ &
   $\rightarrow\mid\leftarrow$ & $-J^{(k-1)}_a/t$ &
   $\rightarrow\mid\leftarrow$ & transition &
   $\rightarrow\mid\leftarrow$ & $-J^{(k)}_a/t$ &
   $\rightarrow\mid$
\\
$G_{1,2}$~~& $\mid\leftarrow$ & & & ${\cal O}(D)e^{-D|t|}$ &
  $\rightarrow\mid\leftarrow$ & transition &
   $\rightarrow\mid\leftarrow$ & $-J^{(k)}_a/t$ & 
   $\rightarrow\mid$
\\
$G_{3}$~~& $\mid\leftarrow$ & & & ${\cal O}(l)e^{-D|t|}$ &
  $\rightarrow\mid\leftarrow$ & transition & 
   $\rightarrow\mid\leftarrow$ & $-J^{(k)}_a/t$ & 
   $\rightarrow\mid$
\\
$b_{1,2}$~~& $\mid\leftarrow$ & & & ${\cal O}(l)+{\cal
  O}(l)e^{-2D|t|}$  
  & $\rightarrow\mid\leftarrow$ & transition &
   $\rightarrow\mid\leftarrow$ & $-J^{(k)}_a/t$ & 
   $\rightarrow\mid$
\\
$b_{3}$~~& $\mid\leftarrow$ & & & $D+{\cal O}(D)e^{-2D|t|}$ &
  $\rightarrow\mid\leftarrow$ & transition &
   $\rightarrow\mid\leftarrow$ & $-J^{(k)}_a/t$ & 
   $\rightarrow\mid$
\\
\hline
\label{k>1table}
\end{tabular}
\begin{quote}
{\bf Table \ref{k>1table}:} The behavior of the elements in the Nahm
matrices for the $k>1$ case. $J_a^{(k)}$ is a $k$-dimensional
irreducible SU(2) representation. The value $-J_a^{(k)}/t$ in
the last five entries indicates that the $P_a$, $G_a$ and $b_a$
together form a pole term $-J_a^{(k)}/t$.
\end{quote}
\end{center} 

{}From Eq.~(\ref{EqnNa}) we see that the variation in the
$N_a$ in the interval $(-t_D,0)$ is negligible. Since the only boundary
values of 
$N_a$ are required to connect to the next interval, this is what
we need to show the equivalence of the Nahm data between the
original and the reduced problem.

Next we turn to the construction equation. Using the notation of
Eq.~(\ref{k>1form}), we have 
\begin{eqnarray}
0&=& -\frac{d}{dt}u + \left[ \left( -N_a + r_a \right) \otimes \sigma_a
\right] u - \left[E_a^\dagger \otimes \sigma_a \right] w - \left
[ F_a^\dagger \otimes \sigma_a \right] z ~,
\label{Eqnua} \\
0&=& -\frac{d}{dt}w - \left[ E_a \otimes \sigma_a \right] u +
\left[ \left( -P_a + r_a \right) \otimes \sigma_a \right] w -
\left[ G_a^\dagger \otimes \sigma_a \right] z ~,
\label{Eqnwa} \\
0&=& -\frac{d}{dt}z - \left[ F_a \otimes \sigma_a \right] u - \left
[ G_a \otimes \sigma_a \right] w + \left[ \left( -b_a + r_a
\right) \otimes \sigma_a \right] z ~.
\label{Eqnza}
\end{eqnarray}

As in Sec.~\ref{Sectsu2}, there are three types of solutions away from
the boundaries. The second and third types of solutions can be ignored
for the same reason as in that section. Here we study in more
detail the first type of solution. In $(-1/D,0)$, these are
normalizable with positive power dependence on $t$, as required in the
$(m+k)$-monopole construction. Using this information and the order of
magnitude of the $G_a$ from Table \ref{k>1table}, we see from the
third term of Eq.~(\ref{Eqnza}) that the $z$ component, which is
exponentially small away from the boundary, can be ${\cal O}(w)$ in
$(-t_D,0)$. The orders of magnitude of the $P_a$ and the $G_a$ can
also cause ${\cal O}(w)$ changes in the $w$ components in
$(-t_D,0)$. However, because of the smaller orders of magnitude of the
$E_a$ and the $F_a$, the $u$ component of this type of solution is
essentially unchanged in $(-t_D,0)$.

So, comparing the original and the reduced problem, we see that the
$u$ component, including its boundary value, is unchanged in the
$D\to \infty$ limit. The $w$ only differ by an amount of ${\cal O}(w)$
within $(-t_D,0)$; this has no effect on the normalization condition,
Eq.~(\ref{su2norm}), and the fields, Eq.~(\ref{su2fields}), in the
limit. The boundary values for $w$ are always zero, as we can see from
the its power dependence on $t$.  The $z$ component is small except
for being ${\cal O}(w)$ in a small region near the boundary, and can 
also be neglected in Eqs.~(\ref{su2norm}) and (\ref{su2fields}).

\end{document}